\documentclass[journal]{IEEEtran}
\IEEEoverridecommandlockouts
\usepackage{balance}
\usepackage{cite}
 \usepackage[keeplastbox]{flushend}
 \usepackage[utf8]{inputenc}
     \usepackage{setspace}
\usepackage{amsmath,amssymb,amsfonts}
\usepackage[compact]{titlesec}         
\titlespacing{\section}{0pt}{0pt}{0pt} 
\usepackage{algorithmic}
\usepackage{graphicx}
\usepackage{subcaption}
\usepackage{epstopdf} 
\usepackage{textcomp}
\usepackage{xcolor}
\newcommand{\RN}[1]{%
\textup{\uppercase\expandafter{\romannumeral#1}}
}
\def\BibTeX{{\rm B\kern-.05em{\sc i\kern-.025em b}\kern-.08em
    T\kern-.1667em\lower.7ex\hbox{E}\kern-.125emX}}
    
\begin{document}
\title{HAPS-assisted Hybrid RF-FSO Multicast Communications: Error and Outage Analysis}
\IEEEoverridecommandlockouts  
\author{Olfa Ben Yahia, \textit{Graduate Student Member}, Eylem Erdogan, \textit{Senior Member}, \textit{IEEE}, Gunes~Karabulut~Kurt, \textit{Senior Member}, \textit{IEEE} %
\thanks{O. Ben Yahia is with the Department of Electronics and Communication Engineering, Istanbul Technical University, Istanbul, Turkey, (e-mail: yahiao17@itu.edu.tr).}%
\thanks{E. Erdogan is with the Department of Electrical and Electronics Engineering, Istanbul Medeniyet University, Istanbul, Turkey, (e-mail: eylem.erdogan@medeniyet.edu.tr). }%
\thanks{G. Karabulut Kurt is with the Poly-Grames Research Center, Department of Electrical Engineering, Polytechnique Montr\'eal, Montr\'eal, Canada (e-mail: gunes.kurt@polymtl.ca). She was with the Department of Electronics and Communication Engineering, Istanbul Technical University when this work was performed.}%
}

\maketitle
\begin{abstract}
In this work, we study the performance of multiple-hop mixed frequency (RF)/free-space optical (FSO) communication-based decode-and-forward protocol for multicast networks. So far, serving a large number of users is considered a promising approach for real-time applications to address the massive data traffic demands. 
 In this regard, we propose two practical use-cases. In the former model, we propose a high altitude platform station (HAPS)-aided mixed RF/FSO/RF communication scheme where a terrestrial ground station intends to communicate with a cluster of nodes through two stratospheric HAPS systems. In the latter model, we assume that the line of sight connectivity is inaccessible between the two HAPS systems due to high attenuation caused by large propagation distances. Thereby, we propose a low Earth orbit satellite-aided mixed RF/FSO/FSO/RF communication. For the proposed scenarios, closed-form expressions of outage probability (OP) and bit error rate are derived. In addition, to illustrate the asymptotic behavior of the proposed models, diversity gains are obtained. Furthermore, ergodic capacity and energy efficiency (EE) are provided for both scenarios. Finally, the simulation results are provided to validate the theoretical derivations. The results show that satellite-aided mixed RF/FSO/FSO/RF scenarios achieve better OP, whereas HAPS-aided mixed RF/FSO/RF scenario can achieve higher EE. 
 
\end{abstract}

\begin{IEEEkeywords}
High altitude platform station, multicast transmission, outage probability, stratospheric attenuation.
\end{IEEEkeywords}

\section{Introduction}
Given the fast growth of real-time applications, high demands for higher throughput, lower latency, and better quality of service, new standardization become essential. In sixth-generation (6G) networks, it is envisioned that data transmission will be conducted over vertical heterogeneous networks, known as VHetNets, which is composed of three layers including space network, aerial network, and terrestrial network \cite{9380673}. Low Earth orbit (LEO) satellites, which are one of the main components of space networks, can provide low round trip delay and a high data rate. In this regard, they become a potent enabler for real-time traffic such as voice and video for global coverage, especially in remote areas, where wired communication is challenging or cannot be employed.

 {Aerial networks can be divided into three groups; low altitude platform stations, unmanned aerial vehicles (UAVs), and high altitude platform stations (HAPS) \cite{9356529}. In particular,} the HAPS systems which are positioned in the stratosphere around 20 km, can combine the characteristics of the terrestrial network (low cost and latency) and of the satellites (wide coverage area). {Recently, HAPS has generated considerable attention in academia and industry communities.} HAPS systems are quasi-stationary aerial vehicles that suffer less from atmospheric turbulence as they are positioned above cloud formations \cite{2010optical}. However, HAPS systems are subject to stratospheric attenuation, which can be caused by sulfuric acid ingredients due to volcanic activity, gases, and polar clouds \cite{giggenbach}.

To successfully fulfill the high data rate requirements, free-space optical (FSO) communication links can be an important enabler for non-terrestrial networks. FSO communication ensures high data rate transmission with a low probability of interception \cite{2018outage}, \cite{erdogan2019performance}. However, FSO communication is subject to atmospheric turbulence and is weather-dependent. Also, the performance of FSO links can be degraded by scintillation, beam spreading, and beam wanders causing significant diffraction effects. More specifically, for the uplink communication, beam wander needs to be considered as it produces a large long-term spot size and produces a significant pointing error that causes serious fluctuations in the transmitted signal. However, in the downlink communication, the effect of beam wander can be neglected, whereas pointing error is still an important drawback as satellites are in motion. In optical communications, aperture averaging, in which a larger aperture diameter is used at the receiver, can be adopted to mitigate the adverse effects of pointing errors and turbulence-induced fading \cite{2018performance}. Furthermore, the optical beam is limited to masking effects over large distances and the line-of-sight (LOS) connectivity can not be established between the satellites and the multicast users due to shadowing and obstacles. Thus, in satellite communication (SatCom), cooperative relaying in which a HAPS can be used as an intermediate relay node can be considered as an effective solution to mitigate the channel impairments {and to reap the benefits offered by both communication systems including higher coverage and lower delay \cite{yahia2021haps}. To enhance the performance of HAPS systems, RF and FSO communications can be used consecutively, which is so-called as mixed RF-FSO communication or hybrid RF/FSO communication \cite{7247760,8064568,yahia2021weather}.}

With the increase of the proliferation of streaming media, video conference, IP-TV, and many other applications of broadcast communication, it is expected that the spectral efficiency of wireless communications can be enhanced by serving a group of users simultaneously \cite{8686225}. Therefore, multicasting is well suitable for these services as it exploits the resources of the network efficiently. According to the international telecommunication union (ITU) recommendations, a multicast group is defined as a set of nodes that have specified their intention to receive packets on a particular multicast group address \cite{ITUR}. Several research works have been conducted to investigate multicasting protocols. {However, the performance analysis of multicast services was mainly limited to terrestrial communication \cite{7521904,5688792,6459483}. For SatCom systems, the performance analysis was performed assuming mostly direct downlink from satellite to the ground \cite{erdogan2020} and focuses on unicast instead of multicast transmission. However, when compared to unicasting, this mode of transmission improves the efficiency of the use of radio resources and is commonly utilized to send the same information to a group of users in any network \cite{ibrahim2019optimizing}. }
{\subsection{Related Work}}
{In prior works, the performance of uplink massive access for mixed RF-FSO
satellite-aerial-terrestrial networks was investigated assuming amplify-and-forward
(AF) protocol \cite{9492801,9314201}. The authors in \cite{9492801} propose an optimization beamforming scheme at the
aerial platform to maximize the minimum average signal-to-interference-plus-noise ratio of the terrestrial users. On the other hand, the authors in \cite{9314201} present a low-complexity and effective beamforming model to implement the uplink space division multiple access for the HAPS-to-satellite channel with an objective
to maximize the ergodic sum rate of the system for the ground users. In \cite{9446153,9319151}, the authors investigate the hybrid RF/FSO for HAPS-based relaying system while considering an integrated space–air-ground network. In both works, the performance analysis was carried out over Gamma-Gamma and shadowed-Rician fading channels in the presence of atmospheric turbulence and attenuation.
The authors in \cite{7521904} introduce a new low-computational radio resource management scheme based on a low complexity greedy algorithm for multicast grouping. 
In \cite{4411865}, the HAPS node is integrated into mobile telecommunications systems leading to enhanced system efficiency in terms of resource capacity, coverage area, and the number of simultaneously active users. Considering geostationary Earth orbit to HAPS architecture, the authors propose a reliable multicast protocol, where the HAPS node ensures the local transmission \cite{4212717}.}

\begin{figure}[!t]
  \centering
    \includegraphics[width=3.2in]{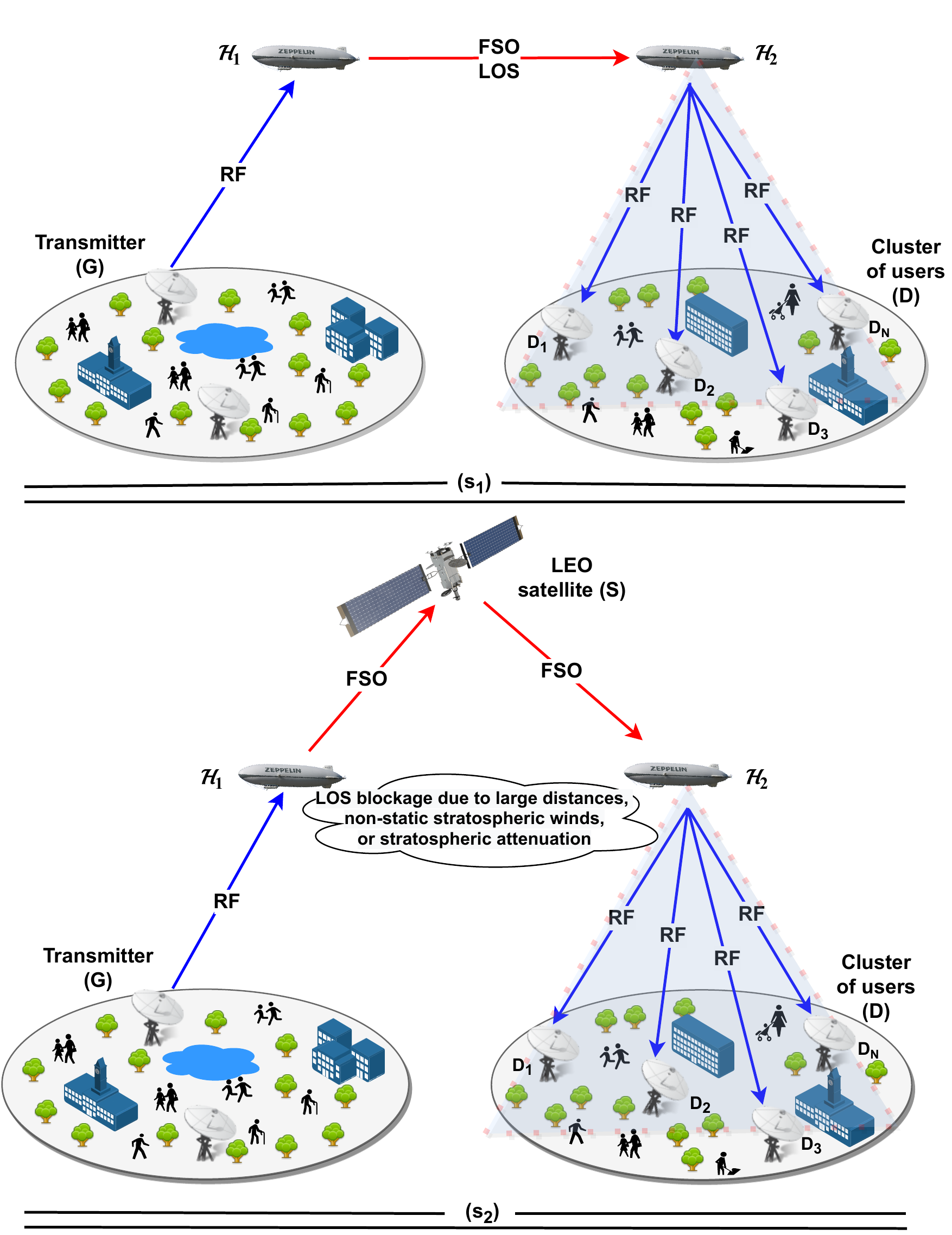}
  \caption{Illustration of the two models. $s_1$) HAPS-aided mixed RF/FSO/RF and $s_2$) Satellite-aided mixed RF/FSO/FSO/RF communication with multicast clustering.}
  \label{fig:model}
\end{figure}
{\subsection{Contributions}}
{To the best of the authors' knowledge, the multicast services in satellite or aerial networks are studied from a network layer perspective. However, the performance analysis of HAPS-aided or satellite-aided-HAPS for multicast services has not been investigated in the literature from a physical layer perspective. Thus, different from the current literature, in this paper, we investigate the physical layer performance of multicast services.} By addressing the above challenges, this paper makes the following contributions:
\begin{itemize}
    \item In this work, two use-cases have been proposed for multicast downlink communication. In the first one, we consider a ground transmitter communicating with a multicast group through the use of two intermediate HAPS systems. In the second use case, the communication is conducted through two HAPS systems connected with an LEO satellite.
   \item For the optical SatCom, we investigate the effect of beam wander-induced pointing error for the uplink, the effect of aperture averaging technique and pointing error for the downlink, and the effect of shadowing severity for RF communication.
    \item We derive closed-form outage probability (OP) {and bit error rate (BER)} expressions for both scenarios by considering stratospheric attenuation, stratospheric turbulence induced-fading for FSO communication, and shadowing and fading for RF communication. {Moreover, to gain further insights and to reveal the asymptotic behavior of the proposed systems, a high signal-to-noise ratio (SNR) analysis is conducted and diversity gain is obtained.}
    {\item We also obtain upper-bounded ergodic capacity expressions and energy efficiency.}
\end{itemize}
\subsection{{Paper Organization}}
The rest of this paper is organized as follows. In Section $\RN{2}$ the signals and system model are presented. {The performance analysis of the system is provided in Section $\RN{3}$.} Numerical results are provided and analyzed in Section $\RN{4}$. {Finally, the conclusion and future directions are drawn in Section $\RN{5}$.}
\vspace{0.2cm}
\section{Signals and System Model}
\vspace{0.2 cm}
In this section, we illustrate the proposed system model of downlink communication for multicast protocol. As shown in Fig. 1, we consider two use-cases. In both cases, we assume the presence of a ground transmitter $G$ that intends to send its information to a set of $N$ ground destinations $D=\left\lbrace D_1, D_2,..., D_N\right\rbrace$ registered under the same multicast address and distributed randomly at hundreds of kilometers far from each other {and connected between them}. More precisely, in the first use-case ($s_1$), the transmission is completed through two HAPS nodes $\mathcal{H}_1$ and $\mathcal{H}_2$ by using optical communication as perfect LOS is established between them. In the second use-case ($s_2$), the LOS connectivity can not be formed between $\mathcal{H}_1$ and $\mathcal{H}_2$ due to the large distance, non-static stratospheric winds, or stratospheric attenuation. Thereby, an LEO satellite-aided optical cooperation is created over three network layers. In particular, $G$ transmits its message to the multicast group in four-time slots over $\mathcal{H}_1$, the LEO satellite $S$, and $\mathcal{H}_2$, all working based on decode and forward (DF) relaying protocol. {DF relaying mode is chosen as it eliminates noise at the relay and the relay can perform perfect detection. In addition, owing to the ease of implementation, DF is frequently used by the researchers.}

{The OP in a conventional multicast network (CVM) is defined in terms of the user with the lowest signal-to-noise ratio (minSNR). While the minSNR policy ensures that all users in a group have the same data rate, it has been shown to be inefficient, especially when a minimum rate or assured quality of service and error-free decoding are required. Specifically, the minSNR technique in CVM clearly allows all users in the group to receive the transmitted data; nevertheless, because it is unable to use multi-user channel diversities to increase network performance, it quickly leads to system saturation \cite{8377988}. So far, there has not been an easy way to use the minSNR strategy in CVM to increase system performance without adding complexity to the optimization process. As a result, we suggest two-phase communication-based cooperative multicast transmission to maximize the system's diversity \cite{6697943}. We also assume that all ground receivers are connected via a terrestrial network. Based on the feedback channel state information (CSI) from the users to $\mathcal{H}_2$, the HAPS node $\mathcal{H}_2$ sends the signal to the user with the best channel characteristics. Finally, the selected user will multicast the data to the users which have worse channel characteristics using the terrestrial network.} In both scenarios, $G$ to $\mathcal{H}_1$ uplink and $\mathcal{H}_2$ to $D$ downlink communications are conducted through RF links that experience the shadowed-Rician fading{, which is very convenient for modeling SatCom links.} The FSO links on the other hand follow the exponentiated Weibull (EW) fading. {EW fading has been found to be the best fit for various aperture sizes for different weather conditions, particularly when aperture averaging is considered to reduce the effects of turbulence and improve the overall performance \cite{erdogan2020}. In addition, we assume that for all FSO links intensity-modulation direct detection (IM/DD) is employed due to its simplicity and low cost.} Moreover, we assume that $\mathcal{H}_2$ is deploying a multi-beam antenna, which makes it able to project multiple spot beams within its potential footprint. {Finally, in the proposed architectures, all links are independent and identically distributed.}
\subsection{HAPS-aided Mixed RF/FSO/RF Communication }
\subsubsection{Ground-to-HAPS Communication }
In the first hop, we assume that RF communication is used due to the destructive effects of aerosols, gases, and atmospheric conditions that highly affect the performance of FSO communication. Thereby, in the first phase, the received signal at $\mathcal{H}_1$ can be given as
\begin{align}
    y_{\mathcal{H}_1}=\sqrt{P_G\mathcal{F}_{\mathcal{H}_1}} h_{G,\mathcal{H}_1} x_{G}+n_{\mathcal{H}_1},
    \label{Eq1}
\end{align}
where $P_G$ denotes the transmit power of $G$, $\mathcal{F}_{\mathcal{H}_1}$ indicates the path loss, $h_{G,\mathcal{H}_1}$ is the channel coefficient between $G$ and $\mathcal{H}_1$, $x_{G}$ represents the transmitted symbol with $\mathbb{E}\left\lbrace|x_{G}|^2\right\rbrace=1$, with $\mathbb{E}\left\lbrace  \cdot \right\rbrace$ is the statistical expectation operator, and $n_{\mathcal{H}_1}$ is the zero-mean complex Gaussian noise at $\mathcal{H}_1$ with noise power spectral density (PSD) $N_{0}$. Based on (1), the instantaneous SNR at $\mathcal{H}_1$ can be written as
\begin{align}
    \gamma_{G,\mathcal{H}_1} = \frac{P_G\mathcal{F}_{\mathcal{H}_1}}{N_{0}}  |h_{G,\mathcal{H}_1}|^2= \overline{\gamma}_{G,\mathcal{H}_1} |h_{G,\mathcal{H}_1}|^2,
    \label{Eq2}
\end{align}
where $\overline{\gamma}_{G,\mathcal{H}_1} =\frac{P_G\mathcal{F}_{\mathcal{H}_1}}{N_{0}} $ is the average SNR of the RF link. Furthermore, the path loss of the RF communication can be given as $\mathcal{F}_{\mathcal{H}_1}[dB]=G_T +G_R-L_F-L_R-L_A$ \cite{9319151},
where $G_T$, $G_R$ denote the transmit and the receiver antenna gains respectively, $L_F$ indicates the free space path loss given as $L_F=92.45+20\log f_r+20 \log L_{G,\mathcal{H}_1}$, $f_r$ represents the frequency in GHz, $L_{G,\mathcal{H}_1}$ represents the propagation distance from $G$ to $\mathcal{H}_1$, $L_R$ indicates the attenuation due to rain in dB/km, and $L_A$ is the gaseous atmosphere loss.

In this work, as we consider shadowed-Rician distribution, the probability density function (PDF) of the instantaneous SNR $\gamma_{G,\mathcal{H}_1} $ can be written as \cite{2019physical}
\begin{align}
f_{{\gamma}_{G,\mathcal{H}_1}}  ({\gamma})&= \sum_{j=0}^{m_{\mathcal{H}_1} -1} \frac{\vartheta_{\mathcal{H}_1}(1-m_{\mathcal{H}_1})_j.(-\delta _{\mathcal{H}_1})^j }{\overline{\gamma} _{G,\mathcal{H}_1}^{j+1} .(j!)^2} \nonumber\\ 
&\times\gamma^j\exp(-\psi_{\mathcal{H}_1} \gamma),
\end{align}
where $m_{\mathcal{H}_1}$ denoting the Nakagami-$m$ fading severity parameter of the corresponding channel, $\vartheta_{G,\mathcal{H}_1}=\frac{1}{2b _{\mathcal{H}_1}}(\frac{2 b_{\mathcal{H}_1} m_{\mathcal{H}_1}}{2 b _{\mathcal{H}_1} m_{\mathcal{H}_1} +\Omega_{\mathcal{H}_1}})^{m_{\mathcal{H}_1}}$ with $2b_{\mathcal{H}_1}$ is the average power of the non-LOS (NLOS) components and $\Omega_{\mathcal{H}_1}$ represents the average power of LOS component.
$\psi_{\mathcal{H}_1}=\frac{\varsigma_{\mathcal{H}_1} - \delta_{G,\mathcal{H}_1}}{\overline{\gamma}_{G,\mathcal{H}_1} }$, $\varsigma_{\mathcal{H}_1}=\frac{1}{2b_{\mathcal{H}_1}}$,   $\delta_{\mathcal{H}_1}=\frac{\Omega_{\mathcal{H}_1}}{2b_{\mathcal{H}_1}(2b_{\mathcal{H}_1} m_{\mathcal{H}_1} +\Omega_{\mathcal{H}_1})}$, and $(\cdot)_j$ is the Pochhammer symbol. In this regard, the cumulative distribution function (CDF) of $\gamma_{G,\mathcal{H}_1} $ is given by
   \begin{align}
       F_{\gamma_{G,\mathcal{H}_1}} (\gamma)&= 1- \sum_{l=0}^{m_{\mathcal{H}_1 -1}} \sum_{q=0}^{l} \frac{\vartheta_{\mathcal{H}_1}(1-m_{\mathcal{H}})_l.(-\delta_{\mathcal{H}_1})^l}{q! (\psi_{\mathcal{H}_1})^{l-q+1}(\overline{\gamma}_{G,\mathcal{H}_1})^{l+1} l!} \nonumber\\
&\times (\gamma)^q\exp(-\psi_{\mathcal{H}_1} \gamma).
 \label{CDF_RF}
   \end{align}
\subsubsection{Inter-HAPS Communication}
Herein, we use FSO communication as LOS conditions can be perfectly satisfied due to the quasi-stationary position of HAPS systems. In the second phase of communication, the received signal at $\mathcal{H}_2$ can be given as
\cite{2018outage}
\begin{align}
    \label{Eq6}
    y_{\mathcal{H}_2}=  \sqrt{\zeta P_{\mathcal{H}_1}} I_{\mathcal{H}_1,\mathcal{H}_2} x_{\mathcal{H}_1} +n_{\mathcal{H}_2},
\end{align}
where $\zeta$ indicates the electrical-to-optical conversion coefficient, $P_{\mathcal{H}_1}$ is the average transmitted optical power of $\mathcal{H}_1$, $x_{\mathcal{H}_1}$ is the transmitted signal of $\mathcal{H}_1$, and $I_{\mathcal{H}_1,\mathcal{H}_2}>0$ represents the received fading gain (irradiance) between the laser of $\mathcal{H}_1$ and the photodetector of $\mathcal{H}_2$ through the optical link. Finally, $n_{\mathcal{H}_2}$ is the is the zero-mean complex Gaussian noise at $\mathcal{H}_2$ with noise PSD $N_{0}$. Thus, the instantaneous SNR at $\mathcal{H}_2$ can be written as
\begin{align}
\label{SNR}
  \gamma_{\mathcal{H}_1,\mathcal{H}_2}=\frac{\zeta {P_{\mathcal{H}_1}}I_{\mathcal{H}_1,\mathcal{H}_2}^2}{N_0} = \overline{\gamma}_{\mathcal{H}_1,\mathcal{H}_2} I_{\mathcal{H}_1,\mathcal{H}_2}^2,
\end{align}
where $\overline{\gamma}_{\mathcal{H}_1,\mathcal{H}_2}$ indicates the average SNR.\\
In FSO communication, the channel gain is considered as a composite optical channel as $I_{\mathcal{H}_1,\mathcal{H}_2}=I^a_{\mathcal{H}_1,\mathcal{H}_2} I^t_{\mathcal{H}_1,\mathcal{H}_2} $, where $I_{\mathcal{H}_1,\mathcal{H}_2}^a$ defines the stratospheric attenuation and $I_{\mathcal{H}_1,\mathcal{H}_2}^t$ models the stratospheric turbulence. The effect of stratospheric turbulence can be modeled by using the EW fading distribution. Thus, the PDF of $I_{\mathcal{H}_1,\mathcal{H}_2}^t$ can be given as
\begin{align}
\label{turbulence}
& f_{I_{\mathcal{H}_1,\mathcal{H}_2}^t}(I)=  \frac{\alpha_{{\mathcal{H}_1},{\mathcal{H}_2}}\beta_{{\mathcal{H}_1},{\mathcal{H}_2}}}{\eta_{{\mathcal{H}_1},{\mathcal{H}_2}}} \left( \frac{I}{\eta_{{\mathcal{H}_1},{\mathcal{H}_2}}}\right) ^{\beta_{{\mathcal{H}_1},{\mathcal{H}_2}}-1}  \nonumber \\
 &\times \exp \left[ -\left( \frac{I}{\eta_{{\mathcal{H}_1},{\mathcal{H}_2}}}\right) ^{\beta_{{\mathcal{H}_1},{\mathcal{H}_2}}}\right]\nonumber \\
 &\times\left( 1-\exp\left[ -\left( \frac{I}{\eta_{{\mathcal{H}_1},{\mathcal{H}_2}}}\right) ^{\beta_{{\mathcal{H}_1},{\mathcal{H}_2}}}\right]\right) ^{\alpha_{{\mathcal{H}_1},{\mathcal{H}_2}}-1},
\end{align}
where $\eta_{{\mathcal{H}_1},{\mathcal{H}_2}}$ is the scale parameter, and $\alpha_{{\mathcal{H}_1},{\mathcal{H}_2}}$, $\beta_{{\mathcal{H}_1},{\mathcal{H}_2}}$ denote the shape parameters which are directly related to the scintillation index. These parameters can be expressed as \cite[Eq. (5)] {yahia2021haps}.
Furthermore, the scintillation index for the horizontal communication can be written as \cite[Sect. (5)]{andrews2005}
     \begin{align}
     \sigma_{I_{{\mathcal{H}_1},{\mathcal{H}_2}}}^2=1.23 C_n^2 K^{7/6} L_{\mathcal{H}_1,\mathcal{H}_2}^{11/7}.
      \end{align} 
In the above equation, $C^2_n$ indicates the refractive index structure parameter \cite[Sect. (5)]{andrews2005}, $K=\frac{2\pi}{\lambda}$ is the optical wave number, $\lambda$ is the optical wavelength, and $L_{\mathcal{H}_1,\mathcal{H}_2}$ represents the propagation distance between $\mathcal{H}_1$ and $\mathcal{H}_2$. 
According to \cite{Fidler}, the communication between two HAPS systems at an altitude of 20 km, is feasible up to 600 km. Herein, it is worth mentioning that the LOS distance might be slightly different from the distance when projecting the HAPS systems on the Earth, due to the curvature nature of the Earth.
For inter-HAPS communication, we consider the effect of stratospheric attenuation, which can be defined as the attenuation of the laser beam caused by molecular absorption, scattering by ice crystals, and some other rare phenomenons such as polar clouds, which are connected to the temperature inside the stratosphere, and stratospheric aerosols that depend on the volcanic activity \cite{giggenbach}. The stratospheric attenuation can be modeled based on the Beer-Lambert law as $I_{\mathcal{H}_1,\mathcal{H}_2}^a= \exp (- \phi L_{\mathcal{H}_1,\mathcal{H}_2})$, where $\phi$ denotes the stratospheric attenuation coefficient \cite{giggenbach}.
\vspace{-0.13cm}
\subsubsection{HAPS-to-Ground Communication}
In the $\mathcal{H}_2$ to $D$ multicast communication, {the optical signal received at $\mathcal{H}_2$ is first decoded then converted into digital RF signal and forwarded to $D$}. Therefore, the received signal at the i-th receiver $D_i$ {providing the best channel characteristics} can be expressed as
\begin{align}
    y_{D_i}=\sqrt{P_{\mathcal{H}_2}\mathcal{F}_{D_i}}h_{\mathcal{H}_2,{D_i}} x_{\mathcal{H}_2} +n_{D_i},
\end{align}
where $P_{\mathcal{H}_2}$ represents the transmit power of $\mathcal{H}_2$, $\mathcal{F}_{D_i}$ indicates the path loss, $h_{\mathcal{H}_2,{D_i}}$ defines the channel information, $x_{\mathcal{H}_2}$ is the transmitted signal from $\mathcal{H}_2$, and $n_{D_i}$ is the Gaussian noise with $N_{0}$ noise PSD. Accordingly, the instantaneous SNR at the multicast destination can be written as
\begin{align}
     \gamma_{\mathcal{H}_2,{D}}=\max ( \gamma_{\mathcal{H}_2,{D_1}}, \gamma_{\mathcal{H}_2,{D_2}},..., \gamma_{\mathcal{H}_2,{D_N}}).
\end{align}
Thus the CDF of $\gamma_{\mathcal{H}_2,{D}}$ becomes
\begin{align}
\label{CDFD}
  F_{\gamma_{\mathcal{H}_2,{D}}}(\gamma) &= \prod_{i=1}^N\Bigg (1- \sum_{p=0}^{m_{D_i} -1} \sum_{k=0}^{p} \frac{\vartheta_{D_i}{(1-m_{D_i})}_p.(-\delta_{D_i})^p}{k! (\psi_{D_i})^{p-k+1}(\overline{\gamma}_{\mathcal{H}_2,D_i})^{p+1} p!} \nonumber\\ 
 & \times (\gamma)^k\exp(-\psi_{D_i} \gamma) \Bigg)  ,
\end{align}
Finally, the end-to-end instantaneous SNR for $s_1$ can be obtained as
\begin{align}
\label{endSNRq1}
    \gamma_{0}^{s_1}=\min \left\lbrace \gamma_{G,\mathcal{H}_1} ; \gamma_{\mathcal{H}_1,\mathcal{H}_2}; \gamma_{\mathcal{H}_2,{D}} \right\rbrace.
\end{align}
\vspace{0.01 cm}
\subsection{Satellite-aided mixed RF/FSO/FSO/RF Communication}
In this subsection, we present the satellite-aided mixed RF/FSO/FSO/RF communication. The $G$ to $\mathcal{H}_1$ and $\mathcal{H}_2$ to $D$ communications were presented in the above section. Thereby, we omit these sections to prevent duplication.
\subsubsection{HAPS-to-Satellite Communication}
In the $\mathcal{H}_1$ to $S$ communication, the received signal can be expressed very similar to (\ref{Eq6}) as $y_S=\sqrt{\zeta P_{\mathcal{H}_1}} I_{\mathcal{H}_1,S} x_{\mathcal{H}_1} +n_{S}$, where $I_{\mathcal{H}_1,S}$ indicates the turbulence induced fading at $S$. The received SNR at $S$ can be written similar to (\ref{SNR}) after changing subscripts ${\mathcal{H}_1,\mathcal{H}_2}$ with ${\mathcal{H}_1,S}$ as $\gamma_{\mathcal{H}_1,S}=\frac{\zeta {P_{\mathcal{H}_1}}I_{\mathcal{H}_1,S}^2}{N_0}$ where  the average SNR can be expressed as
$\overline{\gamma}_{\mathcal{H}_1,S}=\frac{\zeta {P_{\mathcal{H}_1}}}{N_0}$.
Therefore, the CDF of $\gamma_{\mathcal{H}_1,S}$ can be written as
\begin{align}
  \label{CDF_FSO}
 F_{\gamma_{\mathcal{H}_1,S}}(\gamma)&=\sum_{\rho=0}^{\infty} \left( \begin{array}{c} \alpha_{\mathcal{H}_1,S} \\
 \rho
   \end{array}  \right) 
   (-1)^{\rho}\nonumber \\ 
   &\times\exp\left[ -\rho \left( \frac{\gamma}{\eta_{\mathcal{H}_1,S}^2 \overline{\gamma}_{\mathcal{H}_1,S}} \right) ^{\frac{\beta_{\mathcal{H}_1,S}}{2}}\right] , 
   \end{align}
 where $\alpha_{{\mathcal{H}_1},S}$, $\beta_{{\mathcal{H}_1},S}$, and $\eta_{{\mathcal{H}_1},S}$ are the shape parameters and the scale parameter between $\mathcal{H}_1$ and $S$ respectively as defined in \cite[Eq. (5)] {yahia2021haps}{\footnote{The infinite summations in (\ref{CDF_FSO}) require only $5$ terms to converge with a convergence error of $ 2 \times10^{-6}$.}}. In uplink communication, beam wander-induced pointing error effects should be taken into consideration. Thereby, the scintillation index for the uplink communication can be written as \cite[Sect. (12)]{andrews2005}
\begin{align}
\label{EQ:19}
   & \sigma_{I_{\mathcal{H}_1,S}}^2=5.95 ({H_S}-H_{\mathcal{H}_1})^2 \sec^2(\xi_{\mathcal{H}_1,S})\Big( \frac{2W_0}{r_0}\Big)^{\frac{5}{3}}  \Big( \frac{\alpha_{pe}}{W}\Big)^{2} \nonumber\\ 
    & + \exp \left[ \frac{0.49 \sigma_{Bu}^2}{(1+(1.11 +\Theta)\sigma_{Bu}^{\frac{12}{5}})^{\frac{7}{6 }}} + \frac{0.51 \sigma_{Bu}^2}{(1+0.69 \sigma_{Bu}^{\frac{12}{5}})^ {\frac{5}{6}}} \right]-1.
\end{align}
In (\ref{EQ:19}), $H_S$ denotes the altitude of $S$, $H_{\mathcal{H}_1}$ is the altitude of $\mathcal{H}_1$ above the ground level, $\xi_{\mathcal{H}_1,S}$ is the zenith angle between the $\mathcal{H}_1$ and $S$, $W_0$ indicates the beam radius at $\mathcal{H}_1$, and $r_0$ is the Fried's parameter. Furthermore, $W$ indicates the beam size at the receiver, $\alpha_{pe}$ describes the beam wander-induced pointing error, and $\sigma_{Bu}^2$ presents the Rytov variance for uplink communication. The expressions to calculate $\sigma_{I_{\mathcal{H}_1,S}}^2$ are detailed in the Appendix.
\subsubsection{Satellite-to-HAPS Communication}
The received signal at $\mathcal{H}_2$ from the $S$ can be similarly expressed as in (\ref{Eq6}) by just replacing the subscripts as $y_{\mathcal{H}_2}= \sqrt{\zeta P_S} I_{S,\mathcal{H}_2} x_{S} +n_{\mathcal{H}_2}$, whereas  $I_{S,\mathcal{H}_2}=I_{S,\mathcal{H}_2}^a I_{S,\mathcal{H}_2}^t  I^p_{S,\mathcal{H}_2}$ with $I^p_{S,\mathcal{H}_2}$ defines the pointing error component. Thereafter, the instantaneous SNR can be written as given in (\ref{SNR}) by changing the subscripts ${\mathcal{H}_1,\mathcal{H}_2}$ with ${S,\mathcal{H}_2}$. $\alpha_{S,{\mathcal{H}_2}}$, $\beta_{S,{\mathcal{H}_2}}$, $\eta_{S,{\mathcal{H}_2}}$ can be derived using the same equations given in the previous section, and the scintillation index for $S$ to ${\mathcal{H}_2}$ can be written as \cite[Sect. (12)]{andrews2005}. Furthermore, pointing error is considered as one of the major impairments of FSO communication as it involves the displacement of the communicating nodes. In the presence of non-zero boresight pointing errors, the PDF of $I^p_{S,{\mathcal{H}_2}}$ can be expressed as \cite{2014free}
\begin{align}
    f_{I^p_{S,{\mathcal{H}_2}}}(I^p)=\frac{g^2 \exp(\frac{-s^2}{2 \sigma_s^2})}{A_0^{g^2}} (I^p)^{g^2 -1} I_0 \Bigg( \frac{s}{\sigma_s^2} \sqrt{\frac{-w^2_{eq} \ln{\frac{I^p}{A_0}}}{2}} \Bigg),
\end{align}
where $g=w_{eq}/2\sigma_s$ represents the ratio between the equivalent beamwidth $w_{eq}$ and the jitter standard deviation $\sigma_s$, which indicates the severity of the pointing error effects. Furthermore, $s=0$ since we are considering zero boresight error, $A_{0}=[\text{erf}(v)]^2$ denotes the gathered optical power for zero difference between the optical spot center and the detector center, where $v=\sqrt{\pi/2a}/w_{z}$ is the ratio of the aperture radius $a$ and the beamwidth $w_{z}$ at distance $z$ with $w_{z}=\theta z$ with $\theta$ is the transmit divergence angle \cite{2014free}. Finally, $I_0(x)$ defines the modified Bessel function of the first kind with order zero. In the presence of pointing error, the CDF of $\gamma_{S,\mathcal{H}_2}$ becomes \cite{2018performance}
\begin{align}
\label{Perro}
&F_{\gamma_{S,\mathcal{H}_2}}(\gamma)=\frac{\alpha_{S,\mathcal{H}_2} g^2}{\beta_{S,\mathcal{H}_2}} \left( \frac{1}{\eta_{S,\mathcal{H}_2} A_{0}} \sqrt{\frac{\gamma}{\overline{\gamma}_{S,\mathcal{H}_2}}}\right) ^{g^2} \nonumber\\ 
&\times \sum_{i=0}^{\infty} T_{2}(i) G_{2,3}^{2,1} \left( T_{3}(i) \middle \vert \begin{array}{c}
1-T_{1}  , 1 \\
0,1-T_{1} , -T_{1} 
\end{array} 
\right).
\end{align}
In (\ref{Perro}), $G_{p,q}^{m,n} \Big( x\hspace{0.1cm} \Big| \begin{matrix} a_1,...,a_p\\  b_1,...,b_q \end{matrix} \Big)$ denotes the Meijer G-function \cite[eqn. 07.34.02.0001.01]{Wolform}, $T_{1}=g^2/\beta_{S\mathcal{H}_2}$, $T_{2}(i)=(-1)^i \Gamma(\alpha_{S\mathcal{H}_2}) /[i!\Gamma(\alpha_{S\mathcal{H}_2} -i)
(1+i)^{1-T_{1}}]$, and
$T_{3}(i)=(1+i)\left( \frac{1}{\eta_{S\mathcal{H}_2} A_{0}} \sqrt{\frac{\gamma}{\overline{\gamma}_{S\mathcal{H}_2}}}\right) ^{\beta_{S\mathcal{H}_2}}$.{\footnote{The infinite summation in (\ref{Perro}) converges very fast after $5$ iterations with $10^{-6}$ convergence error.}}
In the satellite to HAPS communication, we apply the aperture averaging technique as it can reduce signal fluctuations. Therefore, the aperture size-dependent scintillation index can be expressed as \cite[Eq. (40)]{erdogan2020}.
For $s_2$, the end-to-end instantaneous SNR at the destination $D$ can then be obtained as
\begin{align}
\label{endSNRq2}
    \gamma_{0}^{s_2}=\min \left\lbrace \gamma_{G,\mathcal{H}_1} ; \gamma_{\mathcal{H}_1,S}; \gamma_{S,\mathcal{H}_2}; \gamma_{\mathcal{H}_2,{D}} \right\rbrace,
\end{align} 
where $\gamma_{\mathcal{H}_2,{D}}$ is the instantaneous SNR at the destination which can be written very similar to (\ref{Eq2}).
\vspace{0.1 cm}
{\section{Performance Analysis}
\vspace{0.2 cm}
\subsection{Outage Probability}}
The OP can be described as the probability that the instantaneous SNR $\gamma_0$ falls below a predefined outage threshold $\gamma_{out}$ and it can be written as
\begin{align}
    P_{out}=\Pr[\gamma_0\leq \gamma_{out}].
\end{align}
More precisely, the OP can be derived from the CDF of the end-to-end SNR $\gamma_0$ as $P_{out}=F_{\gamma_0}(\gamma_{out})$. Furthermore, in the case of multi-hop communication, the OP is the probability that the transmitted message is not decoded correctly by at least one of the users at
the end of the communication. 
{For $s_1$, the equivalent CDF of the system SNR ${\gamma_0}$ can be given as
\begin{align}
\label{OP_s1}
    F_{\gamma_0}^{s_1}(\gamma)&=1-  \Big[ \big(  1-F_{\gamma_{G,\mathcal{H}_1}}(\gamma)\big)\big(1-F_{\gamma_{\mathcal{H}_1,\mathcal{H}_2}}(\gamma)\big)  \nonumber\\ 
   &\times\big(1-F_{\gamma_{\mathcal{H}_2,{D}}}(\gamma) \big)
   \Big],
\end{align}
where $F_{\gamma_{\mathcal{H}_1,\mathcal{H}_2}}(\gamma)$ indicates the CDF of $\gamma_{\mathcal{H}_1,\mathcal{H}_2}$. 
For $s_2$, the CDF of the end-to-end SNR can be written as
 \begin{align}
 \label{OP_s2}
   F_{\gamma_0}^{s_2}(\gamma)&=1-  \Big[ \big( 1-F_{\gamma_{G,\mathcal{H}_1}}(\gamma)\big)\big(1-F_{\gamma_{\mathcal{H}_1,S}}(\gamma)\big)  \nonumber\\
   &\times\big(1-F_{\gamma_{S,\mathcal{H}_2}}(\gamma) \big)\big(1-F_{\gamma_{\mathcal{H}_2,{D}}}(\gamma) \big)
   \Big],
\end{align}
and the final expressions of the OP for both scenarios are given at the top of the next page as given (\ref{OP1}) and (\ref{OP2}).}
\begin{figure*} [!h]
\begin{align}
\label{OP1}
 & P_{out}^{s_1} (\gamma_{out}) =  1- \Bigg[ \sum_{l=0}^{m_{\mathcal{H}_1 -1}} \sum_{q=0}^{l} \frac{\vartheta_{\mathcal{H}_1}(1-m_{\mathcal{H}_1})_l.(-\delta_{\mathcal{H}_1})^l}{q! (\psi_{\mathcal{H}_1})^{l-q+1}(\overline{\gamma}_{G,\mathcal{H}_1})^{l+1} l!}
 (\gamma_{out})^q\exp(-\psi_{\mathcal{H}_1} (\gamma_{out})   \times \Bigg( 1- \sum_{\rho=0}^{\infty} \left( \begin{array}{c} \alpha_{\mathcal{H}_1,\mathcal{H}_2} \\
 \rho
   \end{array}  \right)  \nonumber\\ 
   &\times  (-1)^{\rho} \exp\left[ -\rho \left( \frac{\gamma_{out}}{(\eta_{\mathcal{H}_1,\mathcal{H}_2}I_{\mathcal{H}_1,\mathcal{H}_2}^a)^2 \overline{\gamma}_{\mathcal{H}_1,\mathcal{H}_2}} \right) ^{\frac{\beta_{\mathcal{H}_1,\mathcal{H}_2}}{2}}\right] \Bigg) 
   \times \Bigg (1- \prod_{i=1}^N \Bigg (1- \sum_{p=0}^{m_{D_i} -1} \sum_{k=0}^{p} \frac{\vartheta_{D_i}{(1-m_{D_i})}_p.(-\delta_{D_i})^p}{k! (\psi_{D_i})^{p-k+1}(\overline{\gamma}_{\mathcal{H}_2,D_i})^{p+1} p!} \nonumber\\ 
   &\times (\gamma_{out})^k\exp(-\psi_{D_i} \gamma_{out})  \Bigg) \Bigg) \Bigg], 
   \end{align}
\hrulefill
\end{figure*}
\begin{figure*} [!h]
\begin{align}
\label{OP2}
 & P_{out}^{s_2} (\gamma_{out})=  1- \Bigg[ \sum_{l=0}^{m_{\mathcal{H}_1 -1}} \sum_{q=0}^{l} \frac{\vartheta_{\mathcal{H}_1}(1-m_{\mathcal{H}_1})_l.(-\delta_{\mathcal{H}_1})^l}{q! (\psi_{\mathcal{H}_1})^{l-q+1}(\overline{\gamma}_{G,\mathcal{H}_1})^{l+1} l!}
 (\gamma_{out})^q\exp(-\psi_{\mathcal{H}_1} \gamma_{out})\times \Bigg( 1- \sum_{\rho=0}^{\infty} \left( \begin{array}{c} \alpha_{\mathcal{H}_1,S} \\
 \rho
   \end{array}  \right)\nonumber \\ 
   &\times   (-1)^{\rho} \exp\left[ -\rho \left( \frac{\gamma_{out}}{\eta_{\mathcal{H}_1,S}^2 \overline{\gamma}_{\mathcal{H}_1,S}} \right) ^{\frac{\beta_{\mathcal{H}_1,S}}{2}}\right] \Bigg) 
   \times \Bigg( 1- \sum_{\rho_1=0}^{\infty} \left( \begin{array}{c} \alpha_{S,\mathcal{H}_2} \\
 \rho_1
   \end{array}  \right) 
   (-1)^{\rho_1} \exp\left[ -\rho_1 \left( \frac{\gamma_{out}}{(\eta_{S,\mathcal{H}_2}I_{S,\mathcal{H}_2}^a)^2 \overline{\gamma}_{S,\mathcal{H}_2}} \right) ^{\frac{\beta_{S,\mathcal{H}_2}}{2}}\right] \Bigg) \nonumber\\ 
   & \times \Bigg (1- \prod_{i=1}^N \Bigg (1- \sum_{p=0}^{m_{D_i} -1} \sum_{k=0}^{p} \frac{\vartheta_{D_i}{(1-m_{D_i})}_p.(-\delta_{D_i})^p}{k! (\psi_{D_i})^{p-k+1}(\overline{\gamma}_{\mathcal{H}_2,D_i})^{p+1} p!}  (\gamma_{out})^k\exp(-\psi_{D_i} \gamma_{out})  \Bigg) \Bigg) \Bigg].
\end{align}
\hrulefill
\end{figure*}
{\subsection{High SNR Analysis}}
{We now focus on the high SNR evaluation and obtain diversity orders for the proposed models as the obtained exact closed-form expressions provide limited physical insights. Specifically, in the high SNR regime, the OP is given as 
\begin{align}
     P_{out}^{\infty}(\gamma_{out})=(\mathcal{G}_c \overline{\gamma})^{\mathcal{G}_d },
\end{align}
where $\mathcal{G}_c $ presents the coding gain and $\mathcal{G}_d $ indicates the diversity order. The diversity order which defines the slope of the OP versus average SNR curve at asymptotically high SNR, can be calculated as $\mathcal{G}_d=- \lim\limits_{\overline{\gamma} \to\infty} \frac{\log ( P_{out}^{\infty})}{\log(\overline{\gamma})}$.}

{For $s_1$, if the average SNRs of RF links ($\overline{\gamma}_{G,\mathcal{H}_1}$, $\overline{\gamma}_{\mathcal{H}_2,D}$), and the average SNR of the FSO link ($\overline{\gamma}_{\mathcal{H}_1,\mathcal{H}_2}$) go to infinity, therefore, after a few manipulations and by removing the high order terms the 
asymptotic OP can be written as
\begin{align}
      & P_{out}^{s_1,\infty}(\gamma_{out})=\frac{  \vartheta_{\mathcal{H}_1} \gamma_{out}}{\overline{\gamma}_{G,\mathcal{H}_1} }
      + \prod_{i=1}^N \frac{\vartheta_{D_i} \gamma_{out}}{\overline{\gamma}_{\mathcal{H}_2,D_i}}\nonumber \\
     & +  \left( \frac{\gamma_{out}} {(\eta_{\mathcal{H}_1,\mathcal{H}_2} I_{\mathcal{H}_1,\mathcal{H}_2}^a)^2 \overline{\gamma}_{\mathcal{H}_1,\mathcal{H}_2}}\right) ^{\frac{\alpha_{\mathcal{H}_1,\mathcal{H}_2}\beta_{\mathcal{H}_1,\mathcal{H}_2}}{2}} .
\end{align}
\normalsize
and the diversity order can obtained as
\begin{align}
\label{Gdiv}
    \mathcal{G}_d=\min \Big(1,\frac{\alpha_{\mathcal{H}_1,\mathcal{H}_2} \beta_{\mathcal{H}_1,\mathcal{H}_2}}{2},N).
\end{align}
Similarly, for $s_2$, if the average SNRs of RF links ($\overline{\gamma}_{G,\mathcal{H}_1}$, $\overline{\gamma}_{\mathcal{H}_2,D}$), and the average SNR of the FSO links ($\overline{\gamma}_{\mathcal{H}_1,S}$, $\overline{\gamma}_{S,\mathcal{H}_2}$) go to infinity, the OP can be obtained as
\begin{align}
      & P_{out}^{s_2,\infty}(\gamma_{out})=\frac{  \vartheta_{\mathcal{H}_1} \gamma_{out}}{\overline{\gamma}_{G,\mathcal{H}_1} }+ \prod_{i=1}^N \frac{\vartheta_{D_i} \gamma_{out}}{\overline{\gamma}_{\mathcal{H}_2,D_i}} \nonumber \\
      &+\left( \frac{\gamma_{out}} {(\eta_{\mathcal{H}_1,S} I_{\mathcal{H}_1,S}^a)^2 \overline{\gamma}_{\mathcal{H}_1,S}}\right) ^{\frac{\alpha_{\mathcal{H}_1,S}\beta_{\mathcal{H}_1,S}}{2}} \nonumber \\
      &+\left( \frac{\gamma_{out}} {(\eta_{S,\mathcal{H}_2} I_{S,\mathcal{H}_2}^a)^2 \overline{\gamma}_{S,\mathcal{H}_2}}\right) ^{\frac{\alpha_{S,\mathcal{H}_2},\beta_{S,\mathcal{H}_2}}{2}}
     .
\end{align}
\normalsize
Thereafter, the diversity order for $s_2$ can written as
\begin{align}
\label{Gdiv2}
    \mathcal{G}_d=\min \Big(1, \frac{\alpha_{\mathcal{H}_1,S} \beta_{\mathcal{H}_1,S}}{2}, \frac{\alpha_{S,\mathcal{H}_2} \beta_{S,\mathcal{H}_2}}{2}, N).
\end{align} 
It can be noted that the diversity order depends on the turbulence parameters of the FSO link and the number of antennas of the RF links. }
{\subsection{Error Probability Analysis}}
{In this subsection, novel expressions for BER are derived for both scenarios.
For different modulation schemes, the BER can be written as
\begin{align}
\label{BER}
   P_e=\frac{v^u}{2\Gamma(u)} \int_0^\infty \gamma^{u-1} e^{(-v\gamma)}F_{\gamma_0}(\gamma)d\gamma,
\end{align}
where $u$, $v$ present the different binary modulation schemes as given in Table \ref{modulation}. 
\begin{table}[b!]
{\caption{BER parameters for various modulation techniques. }
\small
\label{modulation}
\begin{center}
\begin{tabular}{ |l| l| l| } 
\hline
Binary Modulation & $u$ & $v$ \\
\hline
\hline
Coherent binary frequency shift keying (CBFSK)& $0.5$ & $0.5$\\
Coherent binary phase shift keying (CBPSK) & $0.5$ & $1$\\
Non-coherent binary frequency shift keying (NBFSK)   & $1$ & $0.5$\\ 
Differential binary phase shift keying (DBPSK) & $ 1$ & $1$\\
\hline
\end{tabular}
\end{center}
\label{Tab1}}
\end{table}
\normalsize
\subsubsection{HAPS-aided Mixed RF/FSO/RF Communication}
By substituting (\ref{OP_s1}) in (\ref{BER}) with few manipulations, the BER can be obtained for $s_1$ as 
\begin{align}
\label{Pes1}
    P_e^{s_1}= I_1-I_2+I_3,
\end{align}
The solution of $P_e^{s_1}$ depends on the number of the ground receivers, therefore, for the sake of simplicity, we will consider the presence of two receivers in our derivation that experience the same channel parameters. 
The integral of $I_1$ can be easily obtained as
\begin{align}
\label{I1}
    I_1=\frac{1}{2}.
\end{align}
\begin{figure*} [!ht]
{\begin{align}
\label{I2}
      I_2&= \frac{v^u}{\Gamma(u)} \sum_{l=0}^{m_{\mathcal{H}_1 -1}} \sum_{q=0}^{l} \frac{\vartheta_{\mathcal{H}_1}(1-m_{\mathcal{H}_1})_l.(-\delta_{\mathcal{H}_1})^l}{q! (\psi_{\mathcal{H}_1})^{l-q+1}(\overline{\gamma}_{G,\mathcal{H}_1})^{l+1} l!} \sum_{p=0}^{m_{D_1} -1} \sum_{k=0}^{p} \frac{\vartheta_{D_1}{(1-m_{D_1})}_p.(-\delta_{D_1})^p}{k! (\psi_{D_1})^{p-k+1}(\overline{\gamma}_{\mathcal{H}_2,D_1})^{p+1} p!} \nonumber\\ 
     & \times    \sum_{\rho=1}^{\infty} \binom{\alpha_{\mathcal{H}_1,\mathcal{H}_2}}{\rho} (-1)^{\rho +1}
 \frac{2^{\frac{1}{2}}\beta_{\mathcal{H}_1,\mathcal{H}_2}^{(u+q+k)-\frac{1}{2}} }{(2 \pi)^{\frac{\beta_{\mathcal{H}_1,\mathcal{H}_2}-1}{2}+\frac{1}{2}}} ( v+\psi_{\mathcal{H}_1}+\psi_{D_1})^{-(u+q+k)}\nonumber\\
&\times G_{\beta_{\mathcal{H}_1,\mathcal{H}_2},2}^{2,\beta_{\mathcal{H}_1,\mathcal{H}_2}}\left[ \frac{\Big(\rho \left( \frac{1}{(\eta_{\mathcal{H}_1,\mathcal{H}_2}I_{\mathcal{H}_1,\mathcal{H}_2}^a)^2 \overline{\gamma}_{\mathcal{H}_1,\mathcal{H}_2}} \right) ^{\frac{\beta_{\mathcal{H}_1,\mathcal{H}_2}}{2}}\Big)^2 2^{-2} }{ ( v+\psi_{\mathcal{H}_1}+\psi_{D_1}) ^{{\beta_{\mathcal{H}_1,\mathcal{H}_2}}} \beta_{\mathcal{H}_1,\mathcal{H}_2}^{-\beta_{\mathcal{H}_1,\mathcal{H}_2}}} \middle \vert \begin{array}{c}  \frac{1- (u+q+k)}{\beta_{\mathcal{H}_1,\mathcal{H}_2}},...,\frac{\beta_{\mathcal{H}_1,\mathcal{H}_2}- (u+q+k)}{\beta_{\mathcal{H}_1,\mathcal{H}_2}} \\ 0,\frac{1}{2} \end{array}\right].
\end{align}}
\hrulefill
\end{figure*}
\begin{figure*} [!ht]
{\begin{align}
\label{I3}
 I_3&= \frac{v^u}{2\Gamma(u)} \sum_{l=0}^{m_{\mathcal{H}_1 -1}} \sum_{q=0}^{l} \frac{\vartheta_{\mathcal{H}_1}(1-m_{\mathcal{H}_1})_l.(-\delta_{\mathcal{H}_1})^l}{q! (\psi_{\mathcal{H}_1})^{l-q+1}(\overline{\gamma}_{G,\mathcal{H}_1})^{l+1} l!} \prod_{i=1}^2 \Bigg (\sum_{p=0}^{m_{D_i} -1} \sum_{k=0}^{p}  \frac{\vartheta_{D_i}{(1-m_{D_i})}_p.(-\delta_{D_i})^p}{k! (\psi_{D_i})^{p-k+1}(\overline{\gamma}_{\mathcal{H}_2,D_i})^{p+1} k!}\Bigg) \nonumber\\ 
 &\times\sum_{\rho=1}^{\infty} \binom{\alpha_{\mathcal{H}_1,\mathcal{H}_2}}{\rho} (-1)^{\rho+1}
 (v+\psi_{\mathcal{H}_1}+\psi_{D_1}+\psi_{D_2})^{-(u+q+2k)}\nonumber\\ 
&\times G_{\beta_{\mathcal{H}_1,\mathcal{H}_2},2}^{2,\beta_{\mathcal{H}_1,\mathcal{H}_2}}\left[ \frac{\Big(\rho \left( \frac{1}{(\eta_{\mathcal{H}_1,\mathcal{H}_2}I_{\mathcal{H}_1,\mathcal{H}_2}^a)^2 \overline{\gamma}_{\mathcal{H}_1,\mathcal{H}_2}} \right) ^{\frac{\beta_{\mathcal{H}_1,\mathcal{H}_2}}{2}}\Big)^2 2^{-2} }{ ( v+\psi_{\mathcal{H}_1}+\psi_{D_1}+\psi_{D_2}) ^{{\beta_{\mathcal{H}_1,\mathcal{H}_2}}} \beta_{\mathcal{H}_1,\mathcal{H}_2}^{-\beta_{\mathcal{H}_1,\mathcal{H}_2}}} \middle \vert \begin{array}{c}  \frac{1- (u+q+2k)}{\beta_{\mathcal{H}_1,\mathcal{H}_2}},...,\frac{\beta_{\mathcal{H}_1,\mathcal{H}_2}- (u+q+2k)}{\beta_{\mathcal{H}_1,\mathcal{H}_2}} \\ 0,\frac{1}{2} \end{array}\right].
\end{align}}
\hrulefill
\end{figure*}
$I_2$ and $I_3$ can be derived by first using the following equality $\exp(-bx)=G_{0,1}^{1,0}\left(bx  \middle \vert \begin{array}{c} - \\ 0 \end{array} \right)$ \cite[eqn. (11)]{1990algorithm}. Thereafter, by invoking \cite[eqn. (07.34.21.0013.01)]{Wolform}, $I_2$ and $I_3$ can be obtained in (\ref{I2}) and (\ref{I3}) as given at the top of the next page.
Finally, by substituting (\ref{I1}), (\ref{I2}), and (\ref{I3}) into (\ref{Pes1}), the closed form expression of the BER can be obtained.
\subsubsection{Satellite-aided mixed RF/FSO/FSO/RF Communication}
For $s_2$, the expression of BER can be obtained by substituting (\ref{OP_s2}) in (\ref{BER}) as
\begin{align}
\label{P_s2}
    P_e^{s_2}=L_1 -L_2+L_3.
\end{align}
In (\ref{P_s2}), $L_1=\frac{1}{2}$ and the expressions of $L_2$ and $L_3$ can be given at the top of the next page. Therefore,  by substituting (\ref{L2}) and (\ref{L3}) in (\ref{P_s2}), the closed-form expression of BER for $s_2$ can be obtained.}
\small
\begin{figure*} [!h]
{\begin{align}
\label{L2}
      L_2&= \frac{v^u}{\Gamma(u)} \sum_{l=0}^{m_{\mathcal{H}_1 -1}} \sum_{q=0}^{l} \frac{\vartheta_{\mathcal{H}_1}(1-m_{\mathcal{H}_1})_l.(-\delta_{\mathcal{H}_1})^l}{q! (\psi_{\mathcal{H}_1})^{l-q+1}(\overline{\gamma}_{G,\mathcal{H}_1})^{l+1} l!} \sum_{p=0}^{m_{D_1} -1} \sum_{k=0}^{p} \frac{\vartheta_{D_i}{(1-m_{D_1})}_p.(-\delta_{D_1})^p}{k! (\psi_{D_1})^{p-k+1}(\overline{\gamma}_{\mathcal{H}_2,D_1})^{p+1} p!} \nonumber\\ 
     & \times    \sum_{\rho=1}^{\infty} \sum_{\rho_1=1}^{\infty} \binom{\alpha_{\mathcal{H}_1,S}}{\rho} \binom{\alpha_{S,\mathcal{H}_2}}{\rho_1} (-1)^{\rho +\rho_1+2} 
 \frac{2^{\frac{1}{2}}\beta^{(u+q+k)-\frac{1}{2}} }{(2 \pi)^{\frac{\beta-1}{2}+\frac{1}{2}}} ( v+\psi_{\mathcal{H}_1}+\psi_{D_1})^{-(u+q+k)}\nonumber\\
&\times G_{\beta,2}^{2,\beta}\left[ \frac{ \left(\rho \left( \frac{1}{(\eta_{\mathcal{H}_1,S}I_{\mathcal{H}_1,S}^a)^2 \overline{\gamma}_{\mathcal{H}_1,S}} \right)^{\frac{\beta}{2}} + \rho_1\left( \frac{1}{(\eta_{S,\mathcal{H}_2}I_{S,\mathcal{H}_2}^a)^2 \overline{\gamma}_{S,\mathcal{H}_2}} \right)^{\frac{\beta}{2}}\right)^2 2^{-2} }{ ( v+\psi_{\mathcal{H}_1}+\psi_{D_1}) ^{\beta} \beta^{-\beta}} \middle \vert \begin{array}{c}  \frac{1- (u+q+k)}{\beta},...,\frac{\beta- (u+q+k)}{\beta} \\ 0,\frac{1}{2} \end{array}\right].
\end{align}}
\hrulefill
\end{figure*}
\begin{figure*} [!h]
{\begin{align}
\label{L3}
 L_3&= \frac{v^u}{2\Gamma(u)} \sum_{l=0}^{m_{\mathcal{H}_1 -1}} \sum_{q=0}^{l} \frac{\vartheta_{\mathcal{H}_1}(1-m_{\mathcal{H}_1})_l.(-\delta_{\mathcal{H}_1})^l}{q! (\psi_{\mathcal{H}_1})^{l-q+1}(\overline{\gamma}_{G,\mathcal{H}_1})^{l+1} l!} \prod_{i=1}^2 \Bigg (\sum_{p=0}^{m_{D_i} -1} \sum_{k=0}^{p}  \frac{\vartheta_{D_i}{(1-m_{D_i})}_p.(-\delta_{D_i})^p}{k! (\psi_{D_i})^{p-k+1}(\overline{\gamma}_{\mathcal{H}_2,D_i})^{p+1} k!}\Bigg) \nonumber\\ 
 &\times\sum_{\rho=1}^{\infty} \sum_{\rho_1=1}^{\infty} \binom{\alpha_{\mathcal{H}_1,S}}{\rho} \binom{\alpha_{S,\mathcal{H}_2}}{\rho_1} (-1)^{\rho +\rho_1+2}  \frac{2^{\frac{1}{2}}\beta^{(u+q+2k)-\frac{1}{2}} }{(2 \pi)^{\frac{\beta-1}{2}+\frac{1}{2}}}
 (v+\psi_{\mathcal{H}_1}+\psi_{D_1}+\psi_{D_2})^{-(u+q+2k)}\nonumber\\ 
&\times G_{\beta,2}^{2,\beta}\left[ \frac{ \left(\rho \left( \frac{1}{(\eta_{\mathcal{H}_1,S}I_{\mathcal{H}_1,S}^a)^2 \overline{\gamma}_{\mathcal{H}_1,S}} \right)^{\frac{\beta}{2}} + \rho_1\left( \frac{1}{(\eta_{S,\mathcal{H}_2}I_{S,\mathcal{H}_2}^a)^2 \overline{\gamma}_{S,\mathcal{H}_2}} \right)^{\frac{\beta}{2}}\right)^2 2^{-2} }{ ( v+\psi_{\mathcal{H}_1}+\psi_{D_1}+\psi_{D_2}) ^{\beta} \beta^{-\beta}} \middle \vert \begin{array}{c}  \frac{1- (u+q+2k)}{\beta},...,\frac{\beta- (u+q+2k)}{\beta} \\ 0,\frac{1}{2} \end{array}\right].
\end{align}}
\hrulefill
\end{figure*}
\normalsize
{\subsection{Ergodic Capacity}}
{Ergodic capacity $(C_e)$ is defined as the maximum achievable
capacity of the overall system. Mathematically speaking, it can be written as
\begin{align}
\label{Ce}
    C_{e}=\frac{\log_2(e)}{n} \int_0^\infty \frac{1}{1+\gamma} \overline{F}_{\gamma_0}(\gamma)d\gamma,
\end{align}
where $n$ shows the total transmission time which $3$ for $s_1$ and $4$ for $s_2$ and $\overline{F}_{\gamma_0}(\gamma)=1-{F}_{\gamma_0}(\gamma)$. To the best of the authors' knowledge, the expression of ergodic capacity in closed-form for both scenarios cannot be derived. Therefore, by using the following approximation, the ergodic capacity can be tightly upper bounded
\begin{align}
\label{Ce}
   C_e&= \frac{1}{n}  \mathbb{E} [\log_2 (1+\gamma_{0})] \approx C_e^{up} =\frac{1}{n} \log_2 (1+ \mathbb{E} [\gamma_{0}]).
    \end{align}
  For $s_1$, with the aid of (\ref{endSNRq1}), $C_e^{up,s_1}$ can be expressed as
  \small
    \begin{align}
    \label{Cup}
  C_e^{up, s_1}  &=\frac{1}{3} \log_2 \Bigg(1+ \min \Big( \mathbb{E} [\gamma_{G,\mathcal{H}_1}],\mathbb{E} [\gamma_{\mathcal{H}_1,\mathcal{H}_2}],  \mathbb{E} [\gamma_{\mathcal{H}_2,{D}}]\Big) \Bigg).
\end{align}
\normalsize
In (\ref{Cup}), $  \mathbb{E} [\gamma_{G,\mathcal{H}_1}]$ can be obtained as
\begin{align}
\label{E1}
&  \mathbb{E} [\gamma_{G,\mathcal{H}_1}] = \int_0^\infty (1-F_{\gamma_{G,\mathcal{H}_1}}(\gamma)) d\gamma  \nonumber\\
    &= \sum_{l=0}^{m_{\mathcal{H}_1 -1}} \sum_{q=0}^{l} \frac{\vartheta_{\mathcal{H}_1}(1-m_{\mathcal{H}_1})_l.(-\delta_{\mathcal{H}_1})^l}{q! (\psi_{\mathcal{H}_1})^{l-q+1}(\overline{\gamma}_{G,\mathcal{H}_1})^{l+1} l!} \Gamma [1+q] \psi_{\mathcal{H}_1} ^{-(q+1)},
    \end{align}
  whereas, $\mathbb{E} [\gamma_{\mathcal{H}_1,\mathcal{H}_2}]$ can be obtained as
   \begin{align}
   \label{E2}
  &\mathbb{E} [\gamma_{\mathcal{H}_1,\mathcal{H}_2}] = \int_0^\infty (1-F_{\gamma_{\mathcal{H}_1,\mathcal{H}_2}}(\gamma)) d\gamma\nonumber\\
 &=\sum_{\rho=1}^{\infty} \binom{\alpha_{\mathcal{H}_1,\mathcal{H}_2}}{\rho} (-1)^{\rho+1} \rho^{\frac{2}{\beta_{\mathcal{H}_1,\mathcal{H}_2}}}\nonumber \Gamma\Big[1+ \frac{2}{\beta_{\mathcal{H}_1,\mathcal{H}_2}}\Big]\\
 & \times\Bigg(\frac{1}{(\eta_{\mathcal{H}_1,\mathcal{H}_2}I_{\mathcal{H}_1,\mathcal{H}_2}^a)^2 \overline{\gamma}_{\mathcal{H}_1,\mathcal{H}_2}}\Bigg)^{-1} .
   \end{align}
  The expression of $  \mathbb{E} [\gamma_{\mathcal{H}_2,D}]  $ can be given as 
   \begin{align}
   \label{E3}
  & \mathbb{E} [\gamma_{\mathcal{H}_2,D}] = \max \Big[ \mathbf{E} [\gamma_{\mathcal{H}_2,D_1}], ..., \mathbf{E} [\gamma_{\mathcal{H}_2,D_N}]\Big] .
\end{align}
By substituting (\ref{E1}), (\ref{E2}), and (\ref{E3}) into (\ref{Cup}), the expression for $ C_e^{up, s_1}$ can be obtained.
Similarly, $ C_e^{up, s_2}$ for $s_2$ can be easily obtained by following the same steps, where $ C_e^{up, s_2}$can be given as 
\begin{align}
    \label{Cup2}
  C_e^{up, s_2}  &=\frac{1}{4} \log_2 \Bigg(1+ \min \Big( \mathbb{E} [\gamma_{G,\mathcal{H}_1}],\mathbb{E} [\gamma_{\mathcal{H}_1,S}], \nonumber \\
& \mathbb{E} [\gamma_{S,\mathcal{H}_2}],  \mathbb{E} [\gamma_{\mathcal{H}_2,{D}}]\Big) \Bigg).
\end{align}
}
{\subsection{Energy Efficiency}}
{
Energy efficiency (EE, in bits/Hz/Joule) is defined as the ratio of the spectral efficiency ($C_e$, in bits/s/Hz) to the total power consumption ($P_t$, in Watt). The EE can be expressed as \cite{8292545}
\begin{align}
\label{EE}
    EE= \frac{C_e}{P_t}.
\end{align}
For the first scenario $s_1$, we assume equal power consumption. Therefore, total power consumption can be denoted as $P_t^{s_1}=\frac{1}{3} (P_G+P_{\mathcal{H}_1}+P_{\mathcal{H}_2})$. Note that the coefficient $\frac{1}{3}$ accounts for the fact that the entire communication take place ins three phases. $C_e$ is obtained as in (\ref{Cup}). Therefore, by substituting (\ref{E1}), (\ref{E2}), and (\ref{E3}) into (\ref{Cup}), the expression for spectral efficiency $C_e$ can be obtained so as the expression of EE in (\ref{EE}).
Similarly, the EE for $s_2$ can be easily obtained where the total power is given as $P_t^{s_2}=\frac{1}{4} (P_G+P_{\mathcal{H}_1}+P_S +~P_{\mathcal{H}_2})$.}
\vspace{0.1cm}
\section{Numerical Results and Discussions}
\vspace{0.2cm}
In this section, we investigate the performance of our proposed setup based on the theoretical derivations developed in the previous sections. We first verify the derived expressions with Monte Carlo (MC) simulations. For RF communication, we consider different shadowing severity levels: frequent heavy shadowing ($m=1.0$, $b=0.063$, $\Omega=8.94\times 10^{\text{-4}}$), average shadowing ($m=10$, $b=0.126$, $\Omega=0.835$), and infrequent light shadowing ($m=19$, $b=0.158$, $\Omega=1.29$) \cite{2019physical}. For the first use-case $s_1$, we assume $C_n^2=10^{-18}$ m$^{-2/3}$ and different propagation distances between the two HAPS systems. For the second use-case $s_2$, the LEO satellite is located at 500 km of altitude, both HAPS systems are at altitude of 18 km. Furthermore, the beam radius at $\mathcal{H}_1$ is set to $W_0=2$ cm, $C_0=10^{-18}$ m$^{-2/3}$, $\xi_{\mathcal{H}_1,S}=\xi_{S,\mathcal{H}_2}=70$°, and $u_{S}=u_{\mathcal{H}_2}=65$ m/s. For stratospheric attenuation, we consider the case of molecular absorption at the HAPS node, which is $\phi=10^{-5}$. For all figures, the outage threshold is set to $\gamma_{out}=7$ dB.
\begin{figure}[!t]
  \centering
    \includegraphics[width=3.6in]{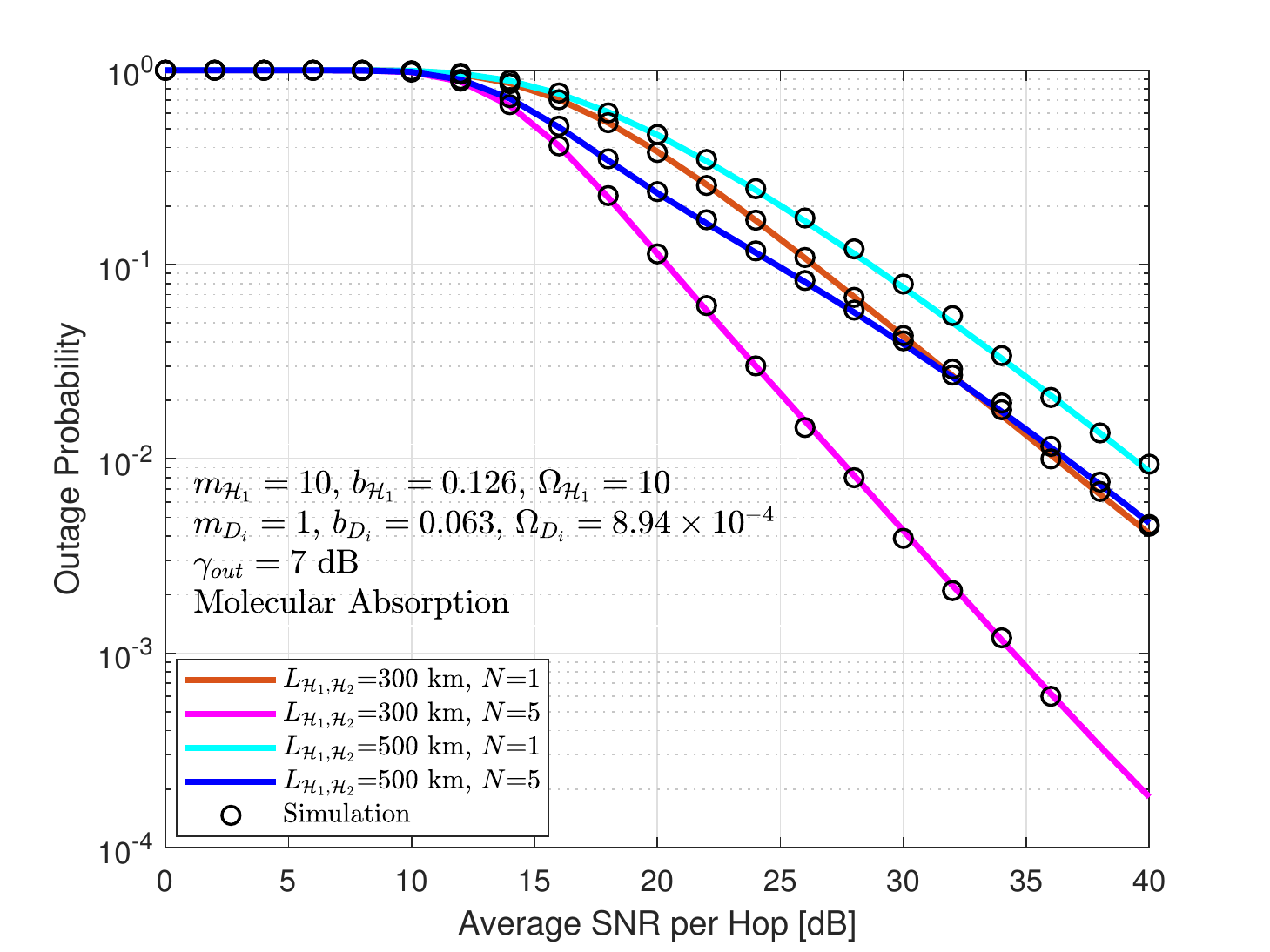}
  \caption{Outage probability performance of $s_1$ for different number of multicast users and HAPS distances $L_{\mathcal{H}_1,\mathcal{H}_2}$.}
  \label{fig:model}
\end{figure}

\begin{figure}[!t]
  \centering
    \includegraphics[width=3.6in]{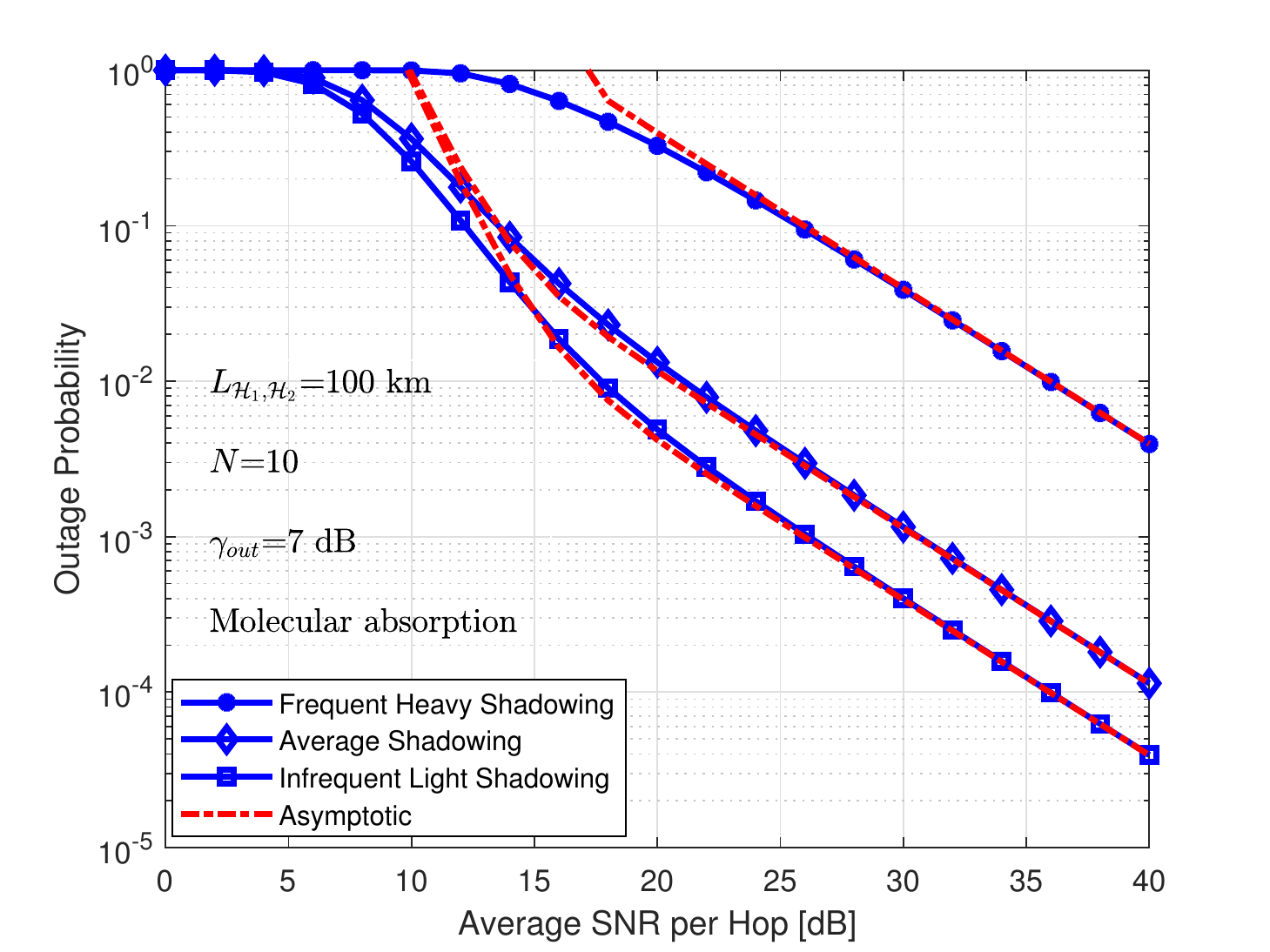}
  \caption{Outage probability performance of $s_1$ for various shadowing levels.}
  \label{fig2l}
\end{figure}

Fig. 2 illustrates the OP as a function of average SNR per hop for a different number of multicast users when the source experiences the average shadowing and the destination experiences the frequent heavy shadowing level. As we can see from the figure, the theoretical results are in good agreement with the MC simulations. In addition, it is obvious that increasing the propagation distance between the HAPS systems deteriorates the overall performance as the scintillation index $\sigma_{I_{\mathcal{H}_1,\mathcal{H}_2}}^2$ increases with the distance from 0.6925 to 1.7667, which means that the turbulence induced-fading can severely affect the communication quality. Furthermore, ected, serving more users at the same time enhances the performance as the number of receivers increases.

In Fig. 3, we evaluate the OP performance of the HAPS-aided scenario for different shadowing levels.
Precisely, we consider a propagation distance of $100$ km between the HAPS systems and a cluster of 10 ground users. Moreover, we assume that the transmitter and the receivers experience the same shadowing level. It is clear from the figure that decreasing the shadowing level improves the overall outage performance. {Finally, we can see from the figure that the asymptotic outage
probability curves almost match the exact outage probability curves for the high SNR region, which validates the obtained expressions.}

Fig. 4 evaluates the OP of the second scenario $s_2$. For the HAPS to satellite ($\mathcal{H}_1$, $S$) and satellite to HAPS communication ($S$, $\mathcal{H}_2$), we consider the same zenith angle and same wind speed. The simulation results showed that the scintillation index is the same for uplink and downlink communications even when considering the presence of beam wander-induced pointing error and this is due to the fact that the HAPS node is located above the clouds' level and it is less sensitive to channel distortion. Furthermore, from this figure, we can observe that the results agree well with MC simulations validating our theoretical expressions. {Similarly, the exact OP results tend to the asymptotic OP results at a high SNR regime, which verifies the correctness of our derivations.} Also, we can see that the OP significantly enhances by increasing the number of downlink receivers. Finally, we observe that increasing the Nakagami-m severity parameter boosts the OP performance.

\begin{figure}[!t]
  \centering
    \includegraphics[width=3.6in]{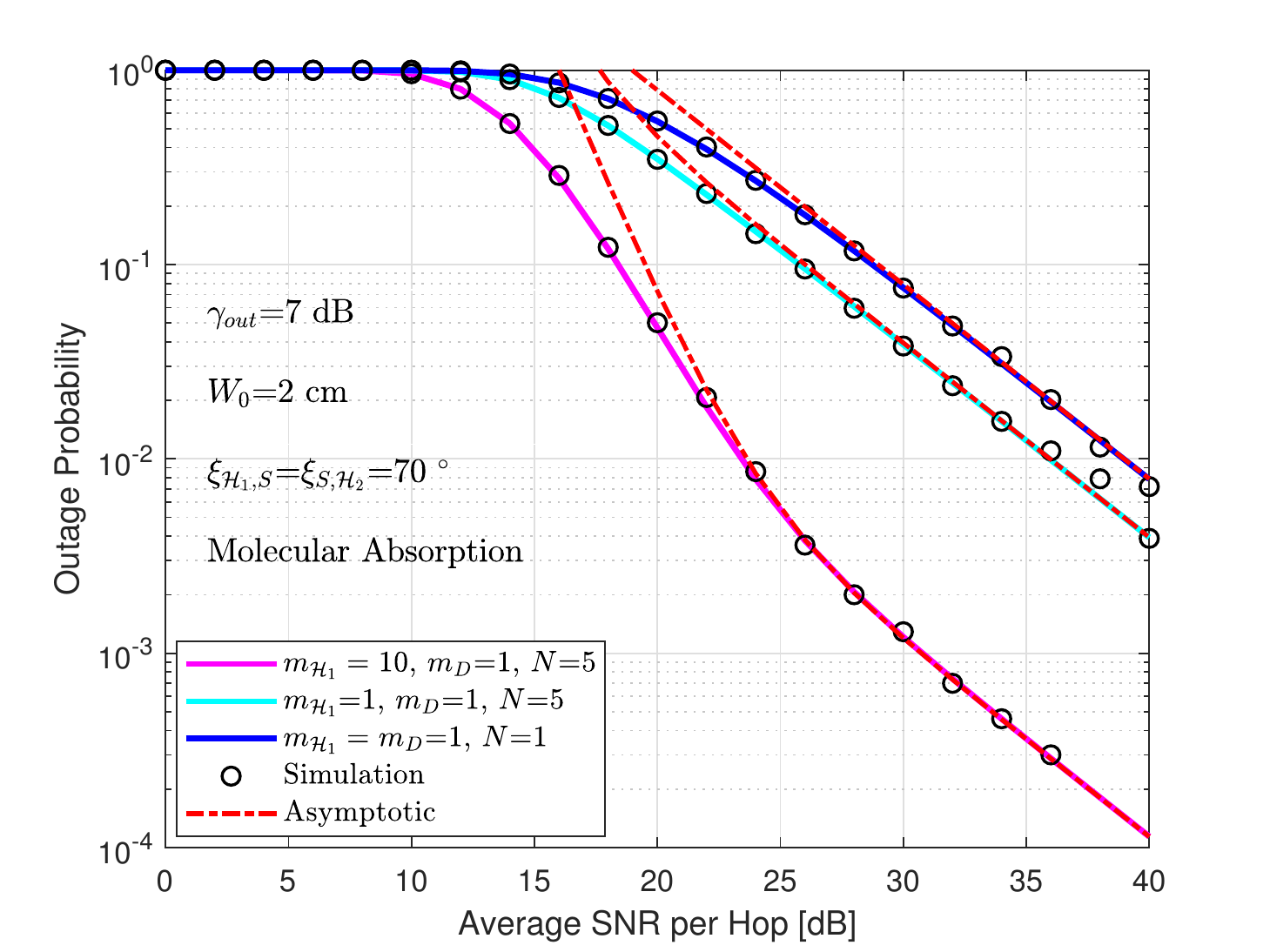}
  \caption{Outage probability performance of $s_2$ for different shadowing levels and various number of multicast users.}
  \label{fig2l}
\end{figure}

\begin{figure}[!t]
  \centering
    \includegraphics[width=3.6in]{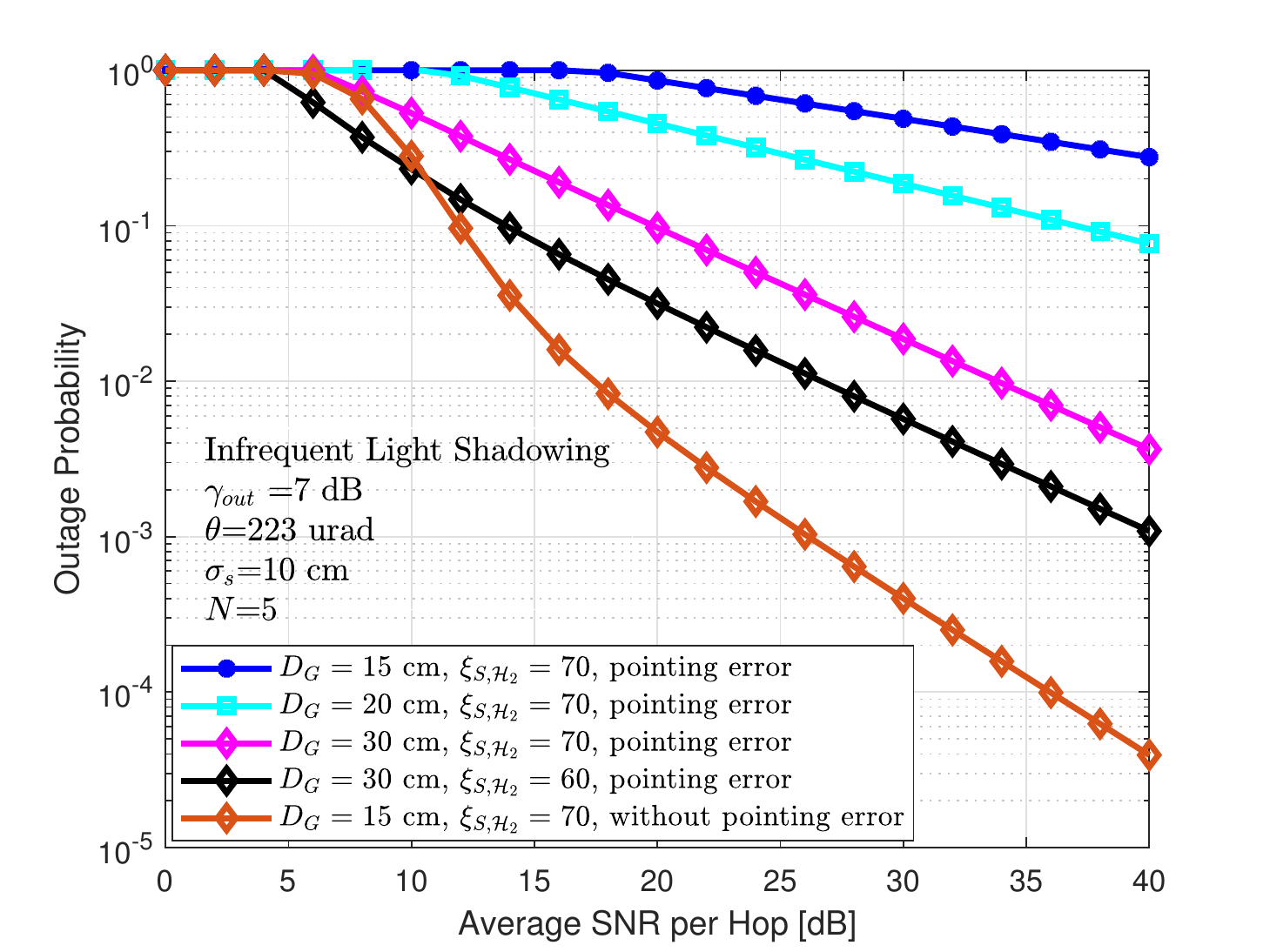}
  \caption{Outage probability performance of $s_2$ for various aperture sizes in the presence of pointing error at $\mathcal{H}_2$.}
  \label{figPE}
\end{figure}

In Fig. \ref{figPE}, we examine the outage performance of $s_2$ by considering the impact of pointing errors in the downlink FSO communication for different aperture sizes. The beam divergence angle is set to $\theta=225$ urad \cite{2017satellite} and the jitter standard deviation is taken as $\sigma_s=10$ cm. As expected, the presence of pointing errors deteriorates the communication and this is due to the misalignment between the satellite and the HAPS $\mathcal{H}_2$. As illustrated in the figure, increasing the aperture size improves the overall performance as it increases the amount of gathered information. Furthermore, choosing lower values of the zenith angle can yield better performance of the proposed scheme.

\begin{figure}[!t]
  \centering
    \includegraphics[width=3.6in]{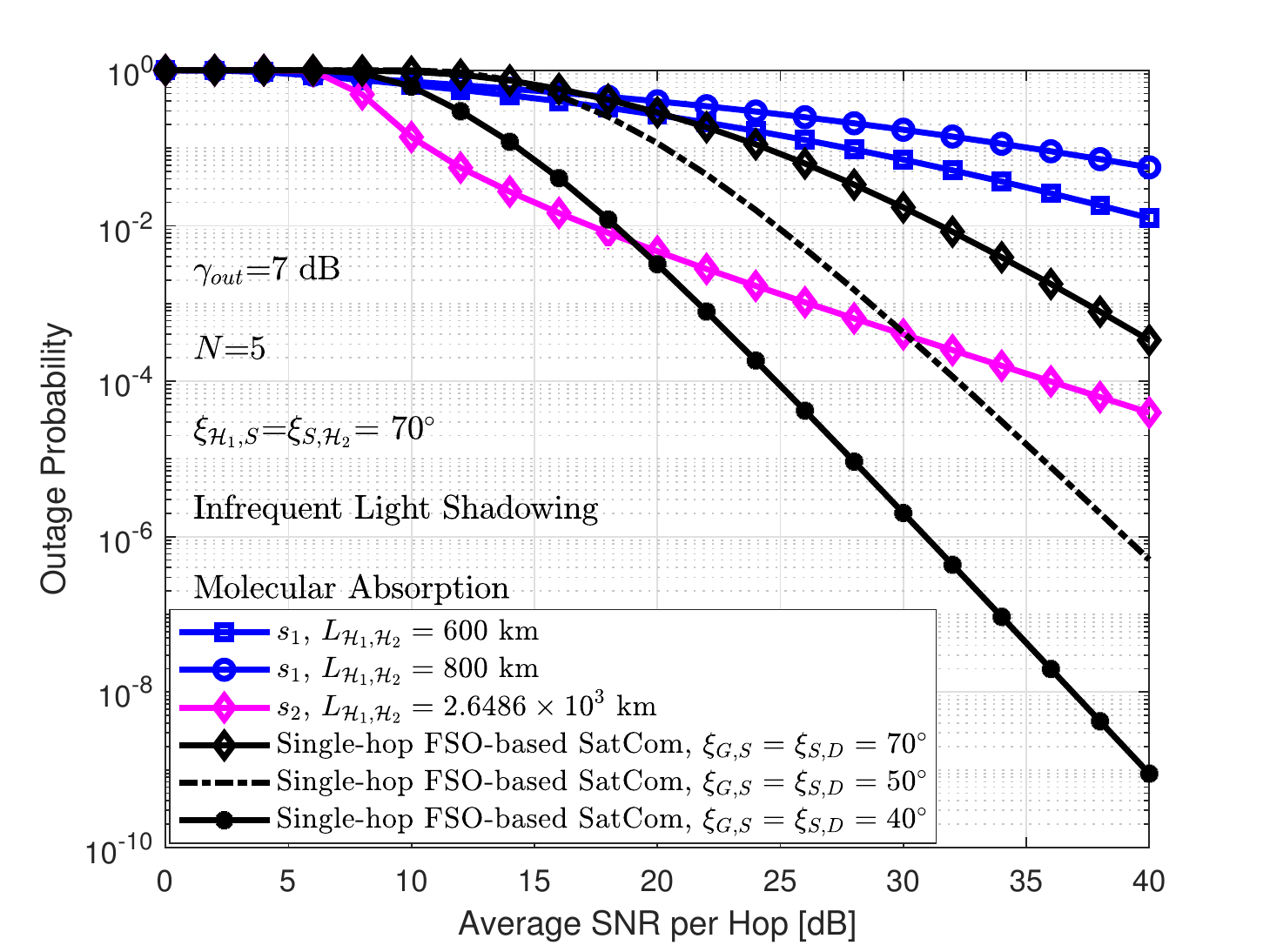}
  \caption{Comparison of $s_1$, $s_2$, and single-hop FSO-based SatCom in terms of outage probability.}
  \label{figCom}
\end{figure}

Fig. \ref{figCom} compares the OP performance of the proposed scenarios {and the OP of single-hop FSO-based SatCom.} For the first scenario, we consider two different propagation distances between $\mathcal{H}_1$ and $\mathcal{H}_2$. For the second scenario, we assume $\xi_{\mathcal{H}_1,S}=\xi_{S,\mathcal{H}_2}=70$° and $u_{S}=u_{\mathcal{H}_2}=65$ m/s, which results in $2.6486 \times 10^3$ km of separation between the two HAPS systems. {For single-hop FSO-based SatCom, we assume thin cirrus cloud condition, for different zenith angles, and $21$ m/s for wind speed. For all scenarios,} we consider five multicast users and that all RF channels experience infrequent light shadowing. It is inferred from the figure that for larger distances between the communicating users, the second scenario is more practical as it performs better even in windy weather and high zenith angles. As expected, at higher propagation distances, the overall performance of the first use-case decreases. {On one hand, we observe that the outage performance of single-hop FSO-based SatCom outperforms $s_1$. However, we expect that for foggy weather $s_1$ could provide better performance than single-hop FSO-based SatCom as FSO communication is highly deteriorated by fog conditions. On the other hand, $s_2$ performs better than single-hop FSO-based SatCom for $\xi_{G,S}=\xi_{S,D}=70^\circ$ as the use of HAPS as a relay node in SatCom enhances the communication even though we consider harsh wind speed level. However, as expected, decreasing the zenith angle for single-hop FSO-based SatCom significantly improves the performance. Specifically, we can see a difference of $14$ dB at $10^{-4}$ between $s_2$ and single-hop FSO-based SatCom for $\xi_{G,S}=\xi_{S,D}=40^\circ$. }

\begin{figure}[!t]
  \centering
    \includegraphics[width=3.6in]{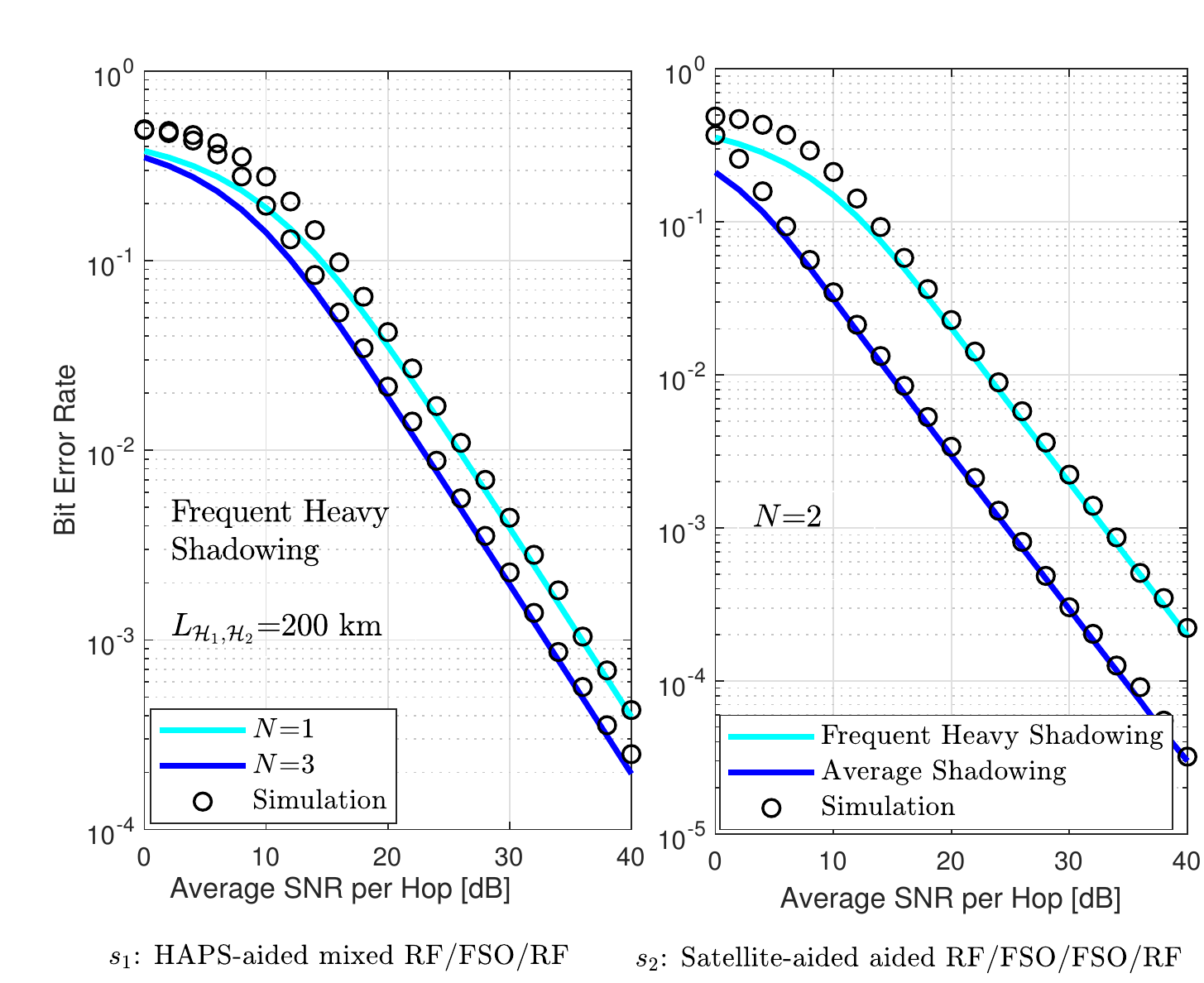}
  \caption{Bit error rate vs. average SNR for $s_1$ and $s_2$.}
  \label{figBER}
\end{figure}
{      
Figure \ref{figBER} depicts the BER vs. average SNR for both scenarios while considering CBPSK modulation as it provides better performance. For $s_1$, we consider the frequent heavy shadowing effect for all RF links and $200$ km that separate $\mathcal{H}_1$ and $\mathcal{H}_2$ for different numbers of the ground receivers. It is evident from the figure that increasing $N$ yields an enhanced performance. Furthermore, as observed, the simulation results confirm our theoretical derivations. Under $s_2$, we assume the presence of two ground receivers while different shadowing levels. Specifically, the simulation results show that $\beta_{\mathcal{H}_1, S}=2.6765$ and $\beta_{S,\mathcal{H}_2}=2.6910$ and this confirms the use of HAPS as a relay node to overcome the effects of beam wander-induced pointing error. Furthermore, as expected, the system yields better performance when considering the average shadowing effect. In addition, all the simulated results confirm the theoretical
outcomes over the considered range of average SNR.
}
\begin{figure}[!t]
  \centering
    \includegraphics[width=3.6in]{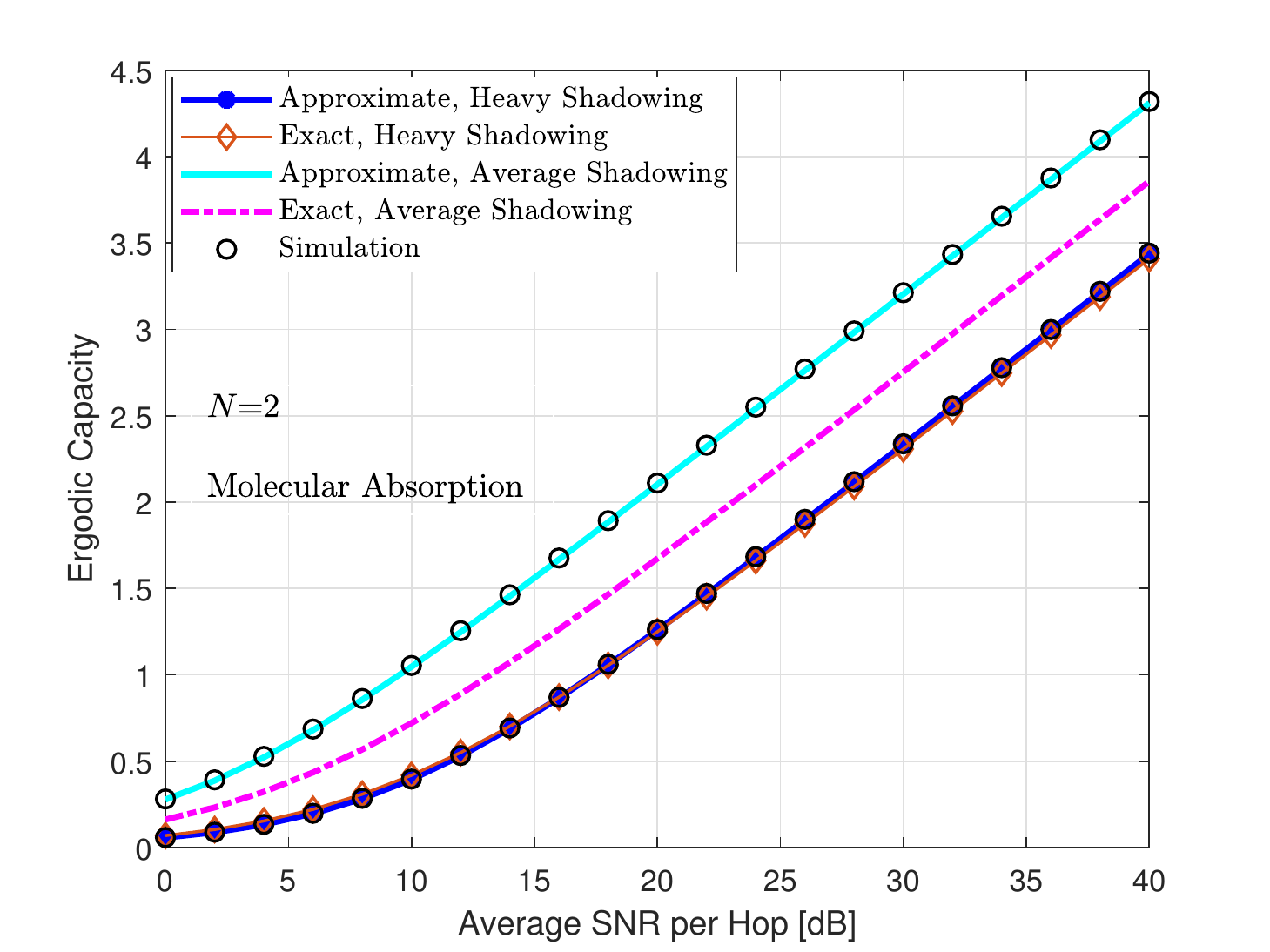}
  \caption{Ergodic capacity performance vs. average SNR for $s_1$.}
  \label{figCap1}
\end{figure}
\begin{figure}[!t]
  \centering
    \includegraphics[width=3.6in]{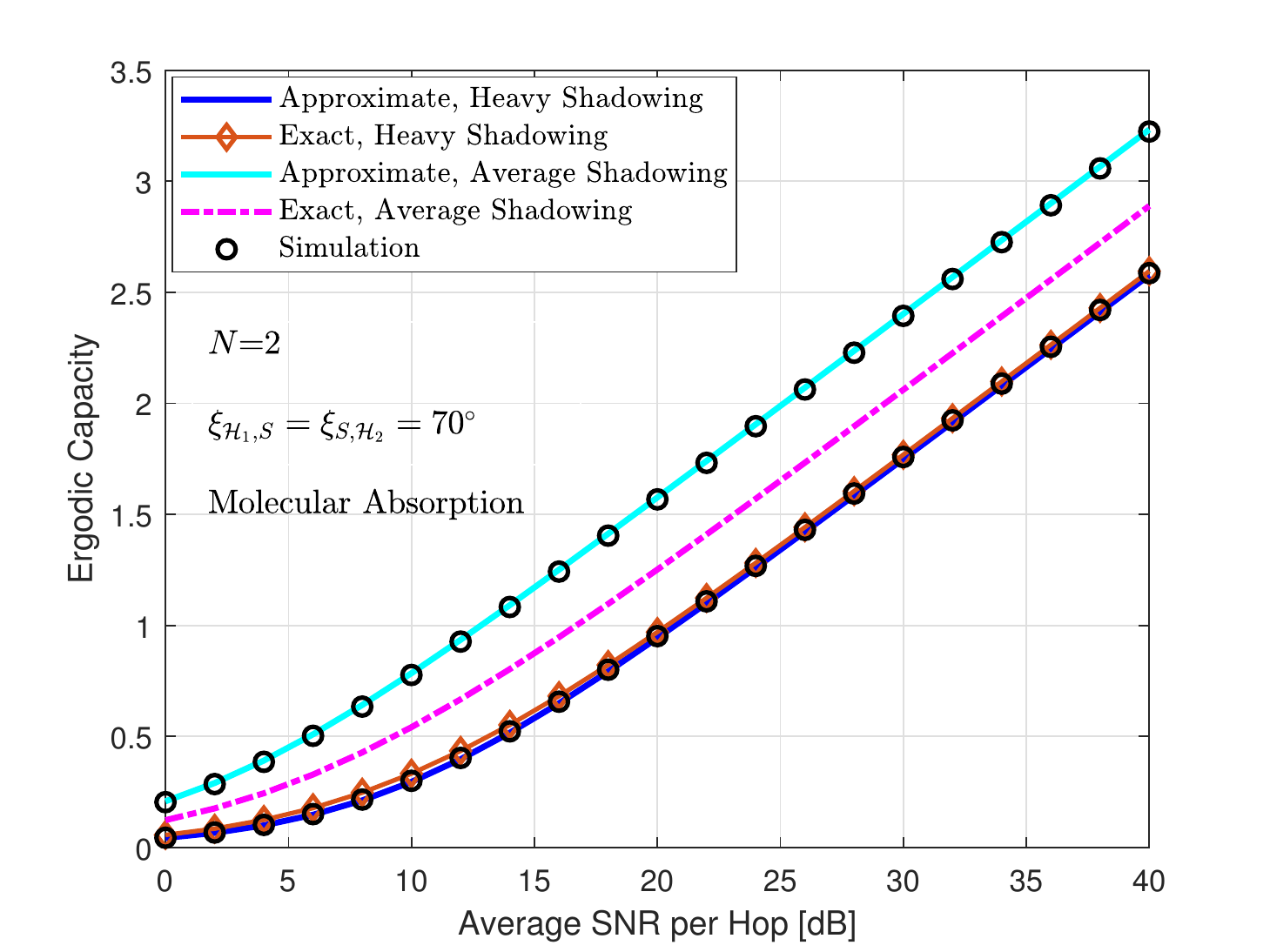}
  \caption{Ergodic capacity performance vs. average SNR for $s_2$.}
  \label{figCap2}
\end{figure}

{Fig. \ref{figCap1} and Fig. \ref{figCap2} show the ergodic capacity performance for $s_1$ and $s_2$ using the upper-bound ergodic capacity and MC simulations for the exact ergodic capacity. As can be seen in the figures, for heavy shadowing, the approximate findings are in good agreement with the exact results. However, for average shadowing, the exact results are upper-bounded with the approximate curves. Furthermore, as expected, the ergodic capacity in average shadowing outperforms the results of heavy shadowing. Additionally, the upper-bound curves agree well with MC simulations confirming the accuracy of the theoretical analysis. It is also inferred from the figures that $s_1$ achieves better capacity than $s_2$. } 

\begin{figure}[!t]
  \centering
    \includegraphics[width=3.6in]{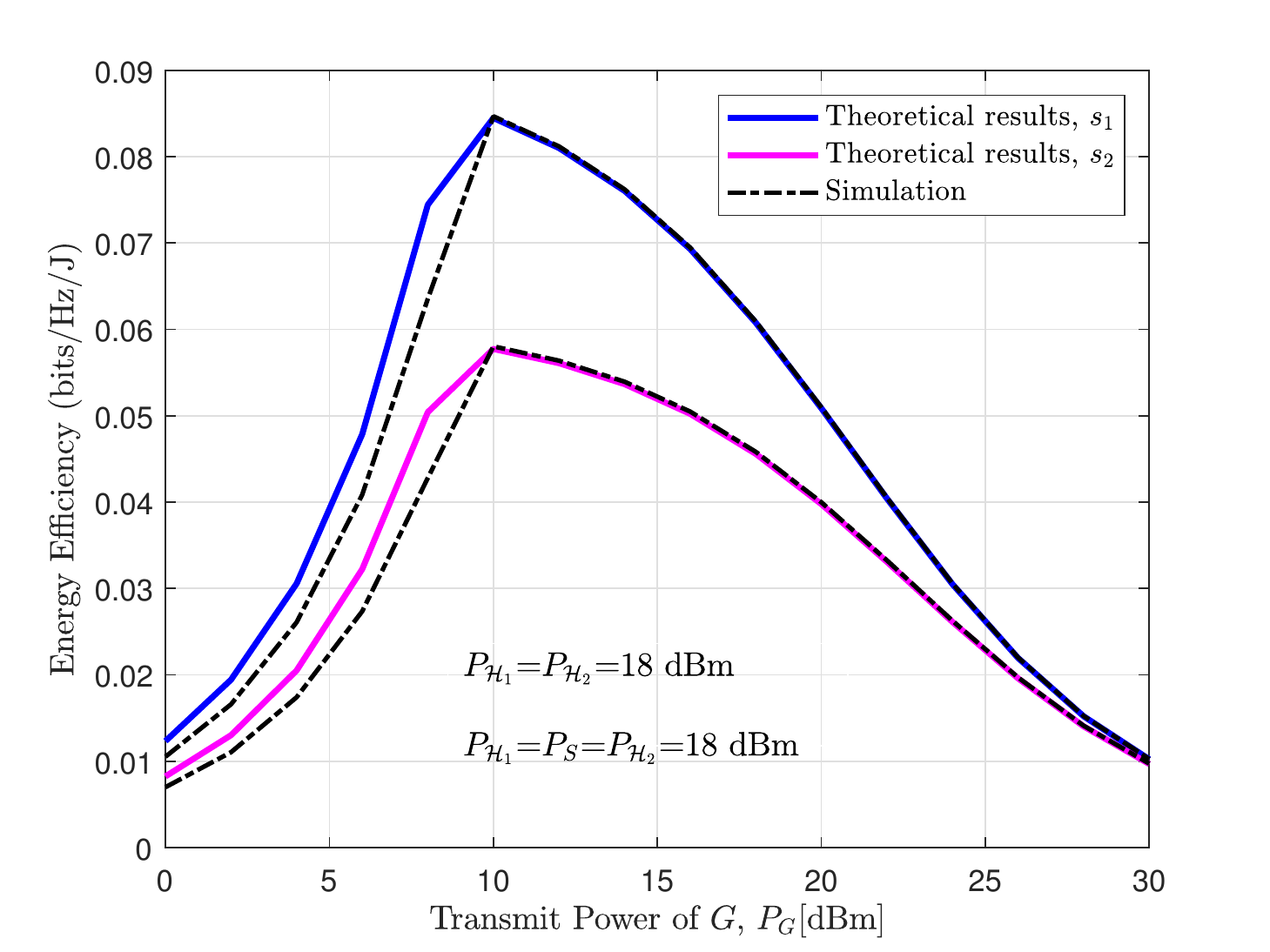}
  \caption{Energy efficiency performance vs. $P_G$ [dBm] for $s_1$ and $s_2$.}
  \label{figEE}
\end{figure}

{Fig. \ref{figEE} illustrates the EE of both scenarios $s_1$ and $s_2$, while assuming the same conditions of shadowing. As can be seen, the EE increases at first and then deteriorates as $P_G$ increases in both scenarios. Moreover, the results have shown that $s_1$ could achieve higher EE than $s_2$. }

Finally, some design guidelines can be summarized as follows:
\begin{itemize}
    \item We observe from the results that the effect of beam wander-induced pointing error from HAPS to satellite is almost negligible. Thus, deploying a HAPS node as a relay between the ground station and the LEO satellite significantly improves the uplink performance.
    \item The results have shown that multicast services can increase the diversity gain and achieves a better outage performance.
    \item The simulations have shown that the use of an LEO satellite can be a promising solution for inter-HAPS communication when HAPS systems are deployed far from each other.
    \item The results have shown that the presence of pointing errors in downlink FSO communication deteriorates the overall performance. However, decreasing the zenith angle or increasing the aperture size of the receiver lens can mitigate the impact of pointing errors as they decrease the impact of stratospheric turbulence.
    {\item As observed $s_2$ could achieve better OP, whereas, $s_1$ achieves higher EE. Thus, there is a trade-off between EE and the OP of the proposed scenarios.}
\end{itemize}

{\section{Conclusion and Future Works }}
\vspace{0.1cm}
In this paper, we proposed two different use-cases for multicast applications namely HAPS-aided mixed RF/FSO/RF and satellite-aided mixed RF/FSO/FSO/RF schemes.
For the proposed setups, OP expressions are derived in closed-form by considering shadowed-Rician fading for RF links and EW fading for FSO communication. The derived expressions are validated with MC simulations. For the proposed models, we investigated the impact of beam wander, pointing errors, zenith angle, and aperture averaging technique for FSO communications and the impact of shadowing level for RF communication. From the results, we observed that the use of HAPS as an intermediate node in uplink SatCom alleviates the impact of beam wander-induced pointing error. In addition, the results showed that for larger distances, the satellite-aided mixed RF/FSO/FSO/RF scenario performed better. The aperture averaging technique can be adopted to minimize the effect of atmospheric turbulence and improve communication in the presence of misalignment errors. {Finally, the results have shown that the satellite-aided mixed RF/FSO/FSO/RF scenario achieves better OP and is able to serve more distant users. However, the HAPS-aided mixed RF/FSO/RF scenario achieves better EE.}
 
{The future work directions can be summarized as follows: Optimal power allocation at the source and the HAPS can be studied to further improve the outage performance. Specifically, maximizing energy efficiency while minimizing the outage probability. In addition, we aim to investigate these schemes while assuming hybrid RF/FSO for ground communication. Finally, the secrecy performance analysis for multicast services will be carried out over generalized FSO and RF distributions to understand the performance of the considered system from a secrecy perspective.}

{\textbf{\appendix}
The Fried's parameter in (\ref{EQ:19}) is given as \cite[Sect. (12)]{andrews2005}
\begin{align}
    r_0=\Bigg[0.42 \sec(\xi_{\mathcal{H}_1,S})K^2 \int_{H_{\mathcal{H}_1}}^{H_S} C_n^2(h)dh\Bigg]^{-3/5}.
\end{align}
In uplink communication, $C_n^2 (h)$ is written as
\begin{align}
 C_n^2(h)&=0.00594\Big(\frac{u_S}{27}\Big)^2 \Big(10^{-5} h\Big)^{10} \exp{\Big(-\frac{h}{1000}}\Big)\nonumber\\
  &+2.7 \times10^{-16} \exp{\Big(-\frac{h}{1500}\Big)}+C_0 \exp{\Big(-\frac{h}{100}\Big)},
  \end{align}
where $u_{S} =\sqrt{v_{S}^2 +30.69 v_{S} +348.91} $ is the root-mean-square (RMS) of the wind speed, $v_{S}$ is the wind speed in m/s,
$C_0$ is the nominal value of $C_n^2$ at the receiver. Generally, $C_n^2$ varies from $10^{-12}$ m$^{-2/3}$ for the strong turbulence to $10^{-17}$ m$^{-2/3}$ for the weak turbulence \cite{2019optical}. In addition, $\alpha_{pe}=\frac{\sigma_{pe}}{L_{\mathcal{H}_1,S}}$ where $L_{\mathcal{H}_1,S}$ denotes the propagation distance from $\mathcal{H}_1$ to $S$, and $\sigma_{pe}^2$ is the beam wander-induced pointing errors variance given as
\begin{align}
\sigma_{pe}^2&=0.54(H_S - H_{\mathcal{H}_1})\sec^2(\xi_{\mathcal{H}_1,S}) \Big(\frac{\lambda}{2W_0}\Big)^2 \Big(\frac{2W_0}{r_0}\Big)^{5/3} \nonumber\\ 
& \times\Big[1- \Big( \frac{{C_r}^2 {W_0}^2/{r_0}^2}{1+{C_r}^2{W_0}^2/{r_0}^2}\Big)^{1/6} \Big],
\end{align}
where $C_r$ is a scaling constant, which is assumed as $\pi$ for $\lambda=1550$ nm, $W=~w_0\sqrt{\Theta_0^2+\Lambda_0^2}$, where $\Lambda_0$ and $\Theta_0$ denotes the beam parameters at the transmitter, which can be expressed as $\Theta_0=~1-L_{\mathcal{H}_1,S}/F_0$ and $\Lambda_0=\frac{2L_{\mathcal{H}_1,S}}{K(w_0)^2}$. $F_0$ is the phase front radius of curvature at the transmitter and $\Theta=\frac{\Theta_0}{\Theta_0^2+\Lambda_0^2}$. Finally, $\sigma_{Bu}^2$ can be expressed as
\begin{align}
   \sigma_{Bu}^2=8.7 \mu_{3u} K^{7/6} ({H_S}-H_{{\mathcal{H}_1}})^{5/6} \sec^{11/6}(\xi_{{\mathcal{H}_1},S}),
\end{align}
where $\mu_{3u}$ is given as
\begin{align}
    \mu_{3u}= \Re \int_{H_{\mathcal{H}_1}}^{H_S} C_n^2(h) [i \varepsilon(1-\varepsilon)]^{5/6} dh,
\end{align}
and $\varepsilon$ is the normalized distance variable given as   $~\varepsilon=~1-~(h-H_{\mathcal{H}_1})~/(H_S-H_{\mathcal{H}_1})~$ for the uplink communication.
}
\bibliographystyle{IEEEtran}
\bibliography{ref}

\end{document}